%
%  Q-LOC. SPIN-ANGULAR MOMENTUM FOR LARGE SPHERES NEAR THE 
%  FUTURE NULL INFINITY 
%

\baselineskip 14pt plus 2pt

\font\llbf=cmbx10 scaled\magstep2
\font\lbf=cmbx10 scaled\magstep1

\def\ni{\noindent}

\def\ba{\bf a}
\def\bb{\bf b}
\def\bc{\bf c}
\def\bd{\bf d}

\def\bi{\bf i}
\def\bj{\bf j}
\def\bk{\bf k}

\def\uA{\underline A \,}      \def\bA{\bf A}
\def\uB{\underline B \,}      \def\bB{\bf B}
      \def\bC{\bf C}
      \def\bD{\bf D}

\def\edth{\hskip 3pt {}^{\prime } \kern -6pt \partial }
\def\thorn{ { \rceil \kern -6pt \supset }}

%%%%%%%%%%%%%%%%%%%%%%%%%%%%%%%%%%%%%%%%%%%%%%%%%%%%%%%%%%%%%%%%%%%%%%%%%%%
%%%%%%%%%%%%%%%%%%%%%%%%%%%%%%%%%%%%%%%%%%%%%%%%%%%%%%%%%%%%%%%%%%%%%%%%%%%

\ni
{\llbf On certain quasi-local spin-angular momentum expressions for large 
spheres near the null infinity}\par
\bigskip
\bigskip
\ni
{\bf L\'aszl\'o B. Szabados}\par
\bigskip
\ni
Research Institute for Particle and Nuclear Physics \par
\ni
H-1525 Budapest 114, P.O.Box 49, Hungary \par
\ni
E-mail: lbszab@rmki.kfki.hu \par
\bigskip
\bigskip

\ni
The recently suggested quasi-local spin-angular momentum expressions, 
based on the Bramson superpotential and on the holomorphic or 
anti-holomorphic spinor fields, are calculated for large spheres near 
the future null infinity of asymptotically flat Einstein--Maxwell 
spacetimes. It is shown that although the expression based on the 
anti-holomorphic spinors is finite and unambiguously defined only in 
the center-of-mass frame (i.e. it diverges in general), the 
corresponding Pauli--Lubanski spin is always finite, free of gauge 
ambiguities, and is built only from the gravitational data. Thus it 
defines a gravitational spin expression at the future null infinity. 
The construction based on the holomorphic spinors diverges in presence 
of outgoing gravitational radiation. For stationary spacetimes both 
constructions reduce to the `standard' expression. 

\bigskip
\bigskip
\ni
{\lbf 1. Introduction}\par
\bigskip
\ni
The energy-momentum and angular momentum {\it density} of matter 
fields is described by their energy-mom\-entum tensor $T^{ab}$, whose 
contraction with a Killing field $K_a$ gives a divergence free current 
$T^{ab}K_b$. The flux of this current on a compact, spacelike 
hypersurface $\Sigma$ with boundary ${\cal S}:=\partial\Sigma$ is 
therefore conserved, i.e. depends only on the boundary ${\cal S}$ and 
independent of the rest of the hypersurface. In Minkowski spacetime 
the independent Killing 1-forms are the translations $K^{\ba}_e:=
\nabla_ex^{\ba}$ and the rotations $K^{\ba\bb}_e:=x^{\ba}\nabla_ex
^{\bb}-x^{\bb}\nabla_ex^{\ba}$, where $x^{\ba}$ are the standard 
Cartesian coordinates, ${\ba},{\bb}=0,...,3$. The corresponding 
conserved quantities are $P^{\ba}_{\cal S}:=\int_\Sigma T^{ab}t_aK
^{\ba}_b{\rm d}\Sigma$ and $J^{\ba\bb}_{\cal S}:=\int_\Sigma T^{ab}t
_aK^{\ba\bb}_b{\rm d}\Sigma$, where $t_a$ is the future directed unit 
timelike normal to $\Sigma$ and ${\rm d}\Sigma$ is the induced volume 
element on $\Sigma$. $P^{\ba}_{\cal S} $ and $J^{\ba\bb}_{\cal S}$ 
can be interpreted as the {\it quasi-local} energy-momentum and angular 
momentum of the matter fields associated with the 2-surface ${\cal S}$, 
respectively. If $\Sigma$ extends to the spacelike infinity ${\it i}^0$ 
or to the future (or past) null infinity ${\cal I}^\pm$ of the Minkowski 
spacetime, then, provided the limits exists (by imposing certain fall-off 
and global integral conditions, e.g. parity conditions at the 
spacelike infinity), then they define the {\it global} energy-momentum 
and angular momentum.

In general relativity the usual fall-off conditions on the 3-metric 
and extrinsic curvature on a spacelike hypersurface ensure that the 
ADM energy and spatial momentum are finite and unambiguously defined 
[1,2]. However, to have finite {\it and} unambiguously defined angular 
momentum and center-of-mass additional conditions, e.g. explicit 
{\it parity conditions} on the three-metric $h_{ab}$ and extrinsic 
curvature $\chi_{ab}$, have to be imposed [2,3]. These are preserved 
by the evolution equations, ensure the functional differentiability 
of the Hamiltonian on the whole phase space, and yield unique Poincare 
structure both for the lapse-shift at infinity and the Hamiltonian. 
At spatial infinity the energy-momentum and relativistic angular 
momentum are the value of the functionally differentiable Hamiltonian, 
which are just the familiar ADM energy-momentum and spatial angular 
momentum, and the center-of-mass of Beig and \'O Murchadha [3]. Thus, 
although there are interesting open issues, e.g. whether the angular 
momentum measured at the spacelike infinity enters the Penrose 
inequality or not [4], the spatial infinity case is well understood.

At null infinity there is a generally accepted definition of the 
energy-momentum, which is the Bondi--Sachs energy-momentum. However, 
on the definition of angular momentum there is no consensus at all, 
and there are various suggestions for that: 
The constructions based on the Komar--Winicour--Tamburino expression 
are intended to be associated with any BMS vector field, and, in the 
special case of the boost-rotation vector fields, they can be 
interpreted as angular momentum [5-7]. The suggestion of Ashtekar and 
Streubel is based on symplectic techniques [8], which turned out later 
to be connected with the Winicour--Tamburino linkages [9]. 
The general form of other expressions are based on charge integrals of 
the curvature [10-16]: In particular, although the suggestion of Penrose 
[12] is based on the solutions of the 2-surface twistor equations and 
the concept of `origin' is largely decoupled from the cuts of ${\cal 
I}^+$, it fits nicely to the symplectic structure of ${\cal I}^+$ 
[17-20]. Other remarkable suggestion in this class is due to Moreschi 
[13-16]. He considers (and, together with Dain, prove [15] the 
existence of) a foliation of ${\cal I}^+$ by special cuts (the `nice 
cuts'), intended to model the `system of rest frames'. Thus he is able 
to define not only a (supertranslation--ambiguity--free) angular 
momentum, but higher order moments of the `gravitational field', too. 
In their classic paper Bergmann and Thomson [21] raises the idea that 
while the energy-momentum of gravity is connected with the spacetime 
diffeomorphisms, the angular momentum should be connected with its 
internal $O(1,3)$-symmetry. Thus the angular momentum should be 
analogous with the spin. This idea was formulated mathematically by 
Bramson [22-24], whose specific angular momentum expression was based 
on the superpotential derived from Hilbert's Lagrangian and the solutions 
of the asymptotic twistor equations. Recently, Katz and Lerer [25] 
could recover the Bondi--Sachs energy-momentum from the standard Noether 
analysis using asymptotically flat backgrounds, and suggested an 
expression for the angular momentum and center-of-mass. Interestingly 
enough, although these suggestions are based on different ideas and yield 
mathematically inequivalent definitions, all these resulting expressions 
bear some resemblance to each other. It is not clear whether the recent 
suggestion by Rizzi [26], based on a special foliation of the spacetime 
near the null infinity and reducing to the ADM angular momentum at the 
spatial infinity [27], also takes such a form.

At the quasi-local level the situation is even worse. Although the 
general framework of how the various quasi-local quantities should look 
like is more or less well understood (see e.g. [28]), only a few 
explicit definitions have been suggested for the energy-momentum {\it 
and} angular momentum. The first such suggestion, due to Penrose [12], 
is, however, not complete: In addition to the (symmetric) kinematical 
twistor ${\tt A}_{\alpha\beta}$ two additional twistors, a Hermitian 
metric twistor ${\tt h}_{\alpha\beta'}$ and the analog of the (skew, 
simple) infinity twistor ${\tt I}_{\alpha\beta}$, would be needed to 
reduce the ten complex components of ${\tt A}_{\alpha\beta}$ (the 
quasi-local quantities) to ten real ones, and to isolate four of them 
as the energy-momentum and six of them as the angular momentum. Although 
at the future null infinity these extra twistors exist and the 
construction works properly, reproducing the Bondi--Sachs 
energy-momentum and gives a definition for the angular momentum, it is 
not clear whether the additional twistors exist and the construction is 
viable quasi-locally or not. 
In principle the Brown--York approach also can yield both energy-momentum 
and angular momentum [29,30]. However, instead of a 4-covariant 
energy-momentum this approach yields separated energy and linear 
momentum, and it is not a priori clear whether they form a single 
Lorentz-covariant object. Furthermore, in lack of a unique prescription 
how the reference configuration should be find (imbedding of the 
2-surface into a 3-plane [29], or into the light cone in Minkowski 
spacetime [31,32], or restrict the imbedding in some other way [33]), 
this construction is not complete either. 
To define Lorentz-covariant energy-momentum and angular momentum 
Ludvigsen and Vickers [34] used the Nester--Witten and the Bramson 
superpotentials, respectively, but their choice for the two spinor 
fields in the superpotential depends on the asymptotic structure of 
the spacetime. Thus their definition is not genuinely quasi-local. 
Genuinely quasi-local, manifestly Lorentz-covariant energy momenta, 
based on the Nester--Witten 2-form and either the holomorphic or the 
anti-holomorphic spinor fields, were suggested by Dougan and Mason 
[35], but they did not give any specific definition for the angular 
momentum. Recently we suggested to complete the Dougan--Mason 
energy-momenta by a spin-angular momentum expression based on Bramson's 
superpotential, but, instead of the Ludvigsen--Vickers prescription 
for the spinor fields, we suggested to use holomorphic or 
anti-holomorphic spinor fields [36]. These have already been studied 
in various situations ({\it pp}-waves [36] and small spheres [37]). 
In the present paper these spin-angular momentum expressions will be 
calculated for large spheres near the future null infinity. 

In [22] Bramson introduced his superpotential as the superpotential 
for the conserved $O(1,3)$-current $C^a_{\ba\bb}$, derived from 
Hilbert's second order Lagrangian (considering that as a function of 
the tetrad field and the $O(1,3)$-connection 1-forms). However, he 
defined $C^a_{\ba\bb}$ as the {\it partial} derivative of Hilbert's 
action with respect to the connection 1-forms, while in gauge theories 
the conserved currents corresponding to the internal gauge invariance 
are the {\it variational} derivatives of the particle action with 
respect to the connection 1-forms. In fact, this variational 
derivative is zero. Thus, in Section 2, first we show that Bramson's 
superpotential can be derived from M\o ller's tetrad action in a 
correct way, and then we review the status of our specific quasi-local 
spin-angular momentum expressions and provide the general formulae for 
the large sphere calculations. In Section 3. we review those structures 
of the future null infinity that we need, especially the BMS translations 
and rotations and their spinor form, and quote the asymptotic solution 
of the Einstein--Maxwell equations from [38]. Then, in Section 4, we 
calculate the anti-holomorphic spin-angular momentum for large spheres 
expanding that as a series of $r^{-1}$. We will see that although this 
is diverging linearly and its finite order part is ambiguous in general, 
the Pauli--Lubanski spin vector built from the anti-holomorphic 
Dougan--Mason energy-momentum and spin-angular momentum is always finite 
and well defined. To demonstrate this, we need the $O(r^{-1})$ accurate 
expansion of the Dougan--Mason energy-momentum. Since, however, this 
calculation has been done only for stationary spacetimes [39] (and 
apparently an $r^{-1}$ order term was overlooked), we have to clarify 
the asymptotic behaviour of the energy-momentum, too. In Section 5 we 
show that in stationary spacetimes the anti-holomorphic construction 
reduces to the `standard' expression. The holomorphic construction, 
calculated in Section 6, is diverging quadratically in presence of 
outgoing gravitational radiation, but gives the same `standard' 
expression for stationary spacetimes. Finally, in the Appendix we 
discuss how the various spinor equations determine special spin frames 
on two-spheres, in particular, on round spheres and on large spheres 
near the future null infinity. We show how the BMS translations and 
rotations can be recovered from them. We found that, in addition to 
the usual representations, they can also be recovered from the 
solutions of the limit of the Dirac--Witten equations on the unit 
sphere cuts of ${\cal I}^+$, too. We treat the 2-surface twistor 
equations also in the traditional way instead of the almost 
exclusively followed conformal method.

Our conventions and notations are mostly those of [40]. In particular, 
we use the abstract index notations, and only the underlined and boldface 
indices take numerical values. The signature is -2, the Riemann- and 
Ricci tensors and the curvature scalar are defined by $-R^a{}_{bcd}X^b
Y^cZ^d:=\nabla_Y(\nabla_ZX^a)-\nabla_Z(\nabla_YX^a)-\nabla_{[Y,Z]}X^a$, 
$R_{ab}:=R^e{}_{aeb}$ and $R:=R_{ab}g^{ab}$, respectively. Thus Einstein's 
equations take the form $R_{ab}-{1\over2}Rg_{ab}=-8\pi GT_{ab}$. 
\bigskip
\bigskip

\ni
{\lbf 2. Bramson's superpotential and quasi-local spin-angular momenta 
for spheres}\par
\bigskip
\ni
Let $\{E^a_{\ba},\vartheta^{\ba}_a\}$, ${\ba}=0,...,3$, be an 
orthonormal dual frame field on an open domain $U\subset M$, and 
$\gamma^{\ba}_{e{\bb}}:=\vartheta^{\ba}_c\nabla_eE^c_{\bb}$, the $O(1,
3)$-connection 1-form on $U$. Then M\o ller's Lagrangian [41] takes the 
form ${\cal L}:={1\over16\pi G}\sqrt{\vert g\vert}(R$ $-2\nabla_a(E^a
_{\ba}\eta^{\ba\bb}E^c_{\bc}\gamma^{\bc}_{c{\bb}}))={1\over16\pi G}
\sqrt{\vert g\vert}(E^a_{\ba}E^b_{\bb}-E^a_{\bb}E^b_{\ba})\gamma^{\ba}
_{a{\bc}}\gamma^{\bc\bb}_b$, where $\eta_{\ba\bb}:={\rm diag}(1,-1,-1,
-1)$. Considering the M\o ller action as a functional both of the tetrad 
field and the connection 1-forms independently, $I_U[E^a_{\ba},\gamma
^{\ba}_{a{\bb}}]:=\int_U{\cal L}{\rm d}^4x$, for the $O(1,3)$ current 
we obtain 

$$
C^a_{\ba\bb}:={4\alpha\over\sqrt{\vert g\vert}}{\delta I_U\over\delta
\gamma^{\ba\bb}_a}=-{\alpha\over4\pi G}\nabla_b\bigl(E^a_{\ba}E^b_{\bb}
-E^a_{\bb}E^b_{\ba}\bigr), \eqno(2.1)
$$
\ni
where $\alpha$ is some normalization constant, to be determined in 
some special situation. $C^a_{\ba\bb}$ is identically conserved, and 
we call the corresponding superpotential 2-form, $W^{\ba\bc}_{ac}:={1
\over4\pi G}\vartheta^{\ba}_{[a}\vartheta^{\bc}_{c]}$, the Bramson 
superpotential. The flux integral of this current for some spacelike 
submanifold $\Sigma$ can be represented by the 2-surface integrals of 
the Bramson superpotential on its boundary ${\cal S}:=\partial\Sigma$: 
$J^{\ba\bb}:=-\alpha\oint_{\cal S}W^{\ba\bb}_{ab}{1\over2}\varepsilon
^{ab}{}_{cd}=-\alpha{1\over2}\varepsilon^{\ba\bb}{}_{\bc\bd}\oint
_{\cal S}W^{\bc\bd}_{cd}$. If $\{E^a_{\ba}\}$ is a 
constant orthonormal (i.e. a Cartesian) frame field in Minkowski 
spacetime, then $J^{\ba\bb}$ is zero. To see this, it is enough to 
recall that in this case $\vartheta^{\ba}_a=\nabla_ax^{\ba}$, the 
gradient of the Cartesian coordinate functions, and hence $W^{\ba\bb}
_{cd}$ is an exact 2-form, whose integral on any closed orientable 
2-surface is vanishing. Therefore, $J^{\ba\bb}$ is a measure of how 
much the frame field $\{E^a_{\ba}\}$ is distorted relative to a 
Cartesian basis in flat spacetime, i.e. a measure of the 
non-integrability of $\{E^a_{\ba}\}$. (The current $\sqrt{\vert g
\vert}C^{\ba}_a:=\delta I_U/\delta E^a_{\ba}$ gives the {\it tensorial} 
energy-momentum expression $t^b{}_a$ for the gravitational `field' 
found in [42,43] (and see also [44]): $C^{\ba}_a=\vartheta^{\ba}_bt^b
{}_a$.)

If ${\cal S}$ is any closed orientable spacelike 2-surface in $M$ 
and $\lambda^{\bA}_A$, ${\bA}=0,1$, is a pair of smooth spinor fields 
on ${\cal S}$ such that they form a normalized spin-frame, i.e. 
$\varepsilon^{AB}\lambda^{\bA}_A\lambda^{\bB}_B=\varepsilon^{\bA\bB}$, 
where $\varepsilon^{\bA\bB}$ is the antisymmetric Levi-Civita symbol, 
then $\vartheta^{\ba}_a:=\sigma^{\ba}_{{\bA}{\bB}'}\lambda^{\bA}_A\bar
\lambda^{{\bB}'}_{A'}$ is an orthonormal 1-form field on ${\cal S}$, 
where $\sigma^{\ba}_{{\bA}{\bB}'}$ are the standard $SL(2,{\bf C})$ 
Pauli matrices. Since the Bramson superpotential depends on $\vartheta
^{\ba}_a$ algebraically, the integral $J^{\ba\bb}$ can be expressed in 
terms of the normalized spin frame fields defined only on ${\cal S}$. 
$J^{\ba\bb}$ is independent of the extension of $\{\lambda^{\bA}_A\}$ 
off the 2-surface ${\cal S}$. Translating the tensor name indices of 
$J^{\ba\bb}$ into spinor name indices, and defining its anti-self-dual 
part by $J^{{\bA}{\bA}'{\bB}{\bB}'}=:\varepsilon^{\bA\bB}\bar J^{{\bA}'
{\bB}'}+\varepsilon^{{\bA}'{\bB}'}J^{\bA\bB}$, we find that 

$$
J^{\bA\bB}={{\rm i}\alpha\over8\pi G}\oint_{\cal S}\lambda^{\bA}_{(A}
\lambda^{\bB}_{B)}\varepsilon_{A'B'}.\eqno(2.2)
$$
\ni
This is precisely the 2-surface integral of the spinorial Bramson 
2-form [22]. Therefore, it has a natural Lagrangian interpretation in 
the sense that it is the spinor form of the superpotential of the 
conserved current derived from M\o ller's tetrad Lagrangian. In the 
present paper by quasi-local energy-momentum we mean an expression 
which is based on the 2-surface integral of the Nester--Witten 2-form: 

$$
P^{{\bA}{\bB}'}:={{\rm i}\over8\pi G}\oint_{\cal S}\bigl(\bar\lambda
^{{\bB}'}_{A'}\nabla_{BB'}\lambda^{\bA}_A-\bar\lambda^{{\bB}'}_{B'}
\nabla_{AA'}\lambda^{\bA}_B\bigr). \eqno(2.3)
$$
\ni
For any pair of spinor fields $\lambda^{\bA}_A$ this defines a 
Lorentzian 4-vector, and it is natural to choose the spinor fields in 
(2.2) and (2.3) to be the same. 
(A more detailed discussion of these issues, and, in particular, the 
connection of these concepts and the current $C^{\ba}_a$ will be 
given elsewhere [45].) 
The quasi-local Pauli--Lubanski spin vector will be defined in the 
standard way by 

$$
S_{\bA{\bA}'}:={1\over2}\varepsilon_{\bA{\bA}'\bB{\bB}'\bC{\bC}'
\bD{\bD}'}P^{\bB{\bB}'}\bigl(\varepsilon^{{\bC}'{\bD}'}J^{\bC\bD}
+\varepsilon^{\bC\bD}\bar J^{{\bC}'{\bD}'}\bigr)={\rm i}\bigl(P_{\bA
{\bB}'}\bar J^{{\bB}'}{}_{{\bA}'}-P_{{\bA}'\bB}J^{\bB}{}_{\bA}\bigr). 
\eqno(2.4)
$$
\ni
(If the quasi-local mass $m$, defined by $m^2:=\varepsilon_{\bA\bB}
\varepsilon_{{\bA}'{\bB}'}P^{{\bA}{\bA}'}P^{{\bB}{\bB}'}$, is not zero, 
then the dimensionally correct definition of the Pauli--Lubanski spin 
is ${1\over m}S_{\bA{\bA}'}$.) To complete the construction of these 
quantities, however, the spin frame field $\lambda^{\bA}_A$ must be 
specified on, and only on  ${\cal S}$. 

In the present paper we will assume that the spin frame is holomorphic, 
$\bar m^e\nabla_e\lambda_A=0$, or anti-holomorphic, $m^e\nabla_e\lambda
_A=0$. Here $m^a$ and $\bar m^a$ are the standard complex null vectors 
tangent to ${\cal S}$ and normalized by $\bar m^am_a=-1$, by means of 
which the metric area-element on ${\cal S}$ is $-{\rm i}m_{[a}\bar m
_{b]}$. In fact, one can show that in the generic case there are two 
holomorphic/anti-holomorphic spinor fields and they can be normalized, 
provided ${\cal S}$ is homeomorphic to $S^2$. Thus $P^{{\bA}{\bB}'}$ 
and $J^{\bA\bB}$ that we will study here are the two energy-momenta of 
Dougan and Mason [35], and the two spin-angular momenta that we 
introduced in [36], respectively. Then the formers can also be written 
as $P^{{\bA}{\bB}'}=-\gamma{{\rm i}\over4\pi G}\oint_{\cal S}(\rho'
\lambda^{\bA}_0\bar\lambda^{{\bB}'}_{0'}+\rho\lambda^{\bA}_1\bar\lambda
^{{\bB}'}_{1'})m_{[a}\bar m_{b]}$, where $\gamma=-1$ for the 
holomorphic, and $\gamma=+1$ 
for the anti-holomorphic spinor fields. In Minkowski spacetime, for 
example, the independent holomorphic spinors are anti-holomorphic too, 
and they are just the constant spinor fields restricted to ${\cal S}$, 
when obviously $P^{{\bA}{\bB}'}=0$ and, as we saw above, $J^{\bA\bB}$ 
is also vanishing [36,37]. If $\varepsilon^A_{\uA}:=\{o^A,\iota^A\}$, 
${\uA}=0,1$, is a standard GHP spin frame adapted to the 2-surface, 
$\varepsilon^{\uA}_A:=\{-\iota_A,o_A\}$ is its dual basis and the 
spinor components with respect to this frame are defined by $\lambda
^{\bA}_A=:\varepsilon^{\uA}_A\lambda^{\bA}_{\uA}$, then the condition 
of holomorphy can be written as $\edth'\lambda^{\bA}_1+\sigma'\lambda
^{\bA}_0=0$ and $\edth'\lambda^{\bA}_0+\rho\lambda^{\bA}_1=0$; and the 
condition of anti-holomorphy is equivalent to $\edth\lambda^{\bA}_0+
\sigma\lambda^{\bA}_1=0$ and $\edth\lambda^{\bA}_1 +\rho'\lambda^{\bA}
_0=0$. Thus boldface capital indices are referring to a basis in the 
space of solutions, while the underlined capital indices to the GHP 
spin frame. For example, for a round sphere [39] of radius $r$ the 
two linearly independent anti-holomorphic and holomorphic spinor 
fields are given by (A.2.2) and (A.3.2) of the Appendix, respectively. 
Substituting them into (2.2) we get zero, as could be expected in a 
spherically symmetric system. 

It is known that the anti-holomorphic Dougan--Mason energy-momentum is 
a future directed non-spacelike vector if ${\cal S}$ is the boundary 
of some compact spacelike hypersurface $\Sigma$, the matter fields 
satisfy the dominant energy condition on $\Sigma$, and ${\cal S}$ is 
convex in the weak sense that $\rho'\geq0$ [35]. (For the holomorphic 
construction the analogous convexity condition is $\rho\leq0$.) 
Furthermore, if the dominant energy condition holds on the whole 
domain of dependence $D(\Sigma)$ of $\Sigma$ then $P^{\bA{\bB}'}$ is 
vanishing if and only if $D(\Sigma)$ is flat, and this is also 
equivalent to the existence of two constant spinor fields on ${\cal 
S}$ with respect to the covariant derivative $\Delta_e$ (see the 
Appendix); and $P^{\bA{\bB}'}$ is null if and only if $D(\Sigma)$ has 
a {\it pp}-wave metric and the matter is pure radiation, which is also 
equivalent to the existence of one constant spinor field on ${\cal 
S}$ [46,47]. $J^{\bA\bB}$ has already been calculated for 
(axi-symmetric) ${\cal S}$ bounding a {\it pp}-wave Cauchy development 
and it was shown that the Pauli--Lubanski spin $S_{\bA{\bB}'}$ is 
proportional to the (null) $P_{\bA{\bB}'}$ [36]. 

We have already calculated (2.2) for the Ludvigsen--Vickers-, the 
holomorphic- and the anti-holomorphic spinors for small spheres ${\cal 
S}_r$ of radius $r$ with respect to an observer $t^a$ at a point $o\in 
M$ [37]. Considering this to be a function of the radius, $J_r^{\bA
\bB}$, it can be expanded as a power series of $r$. The leading term 
is $-{4\alpha\over3}\pi r^4T_{AA'BB'}t^{AA'}t^{B'E}\varepsilon^{BF}
{\cal E}^{({\bA}}_E{\cal E}^{{\bB})}_F$ for the Ludvigsen--Vickers 
and the holomorphic spinors, whilst that is its negative for the 
anti-holomorphic spinors. In vacuum the leading term is $-{4\alpha\over
45G}r^6T_{AA'BB'CC'DD'}t^{AA'}t^{BB'}$ $t^{CC'}t^{D'E}\varepsilon^{DF}
{\cal E}^{({\bA}}_E{\cal E}^{{\bB})}_F$ in all cases, where $T_{abcd}$ 
is the Bel--Robinson tensor and $\{ {\cal E}^A_{\bA}, {\cal E}^{\bA}
_A\}$ is the dual Cartesian spin frame at $o$; i.e. $E^a_{\ba}=\sigma
^{\bA{\bA}'}_{\ba}{\cal E}^A_{\bA}\bar{\cal E}^{A'}_{{\bA}'}$ and 
$\vartheta^{\ba}_a=\sigma_{\bA{\bA}'}^{\ba}{\cal E}^{\bA}_A\bar{\cal 
E}^{{\bA}'}_{A'}$ form an orthonormal dual frame at $o$. Thus the `pure 
gravitational field' itself does not seem to contribute to the 
spin-angular momentum in $r^4$ order. On the other hand, for the 
leading term in the similar expansion of the quasi-local 
(anti-self-dual) angular momentum of the matter fields in Minkowski 
spacetime we get ${4\over3}\pi r^4T_{AA'BB'}t^{AA'}t^{B'E}\varepsilon
^{BF}{\cal E}^{({\bA}}_E{\cal E}^{{\bB})}_F$. The coefficient  $rt^{A'E}
\varepsilon^{AF}{\cal E}^{\bA}_{(E}{\cal E}^{\bB}_{F)}$ is just an 
average of the approximate (and in Minkowski spacetime the exact) 
anti-self-dual boost-rotation Killing vector that vanishes at $o$, 
where the average is taken on the unit sphere ${\cal S}$. Therefore, 
identifying this a.s.d. angular momentum of the matter fields with the 
$r^4$ order {\it holomorphic} $J^{\bA\bB}_r$, we obtain $\alpha=-1$, 
but identifying with the {\it anti-holomorphic} $J^{\bA\bB}_r$ we get 
$\alpha=+1$. Thus there are two equally reasonable conventions: 
Requiring the $r^4$ order terms to coincide with that of the angular 
momentum of the matter fields, whenever the sign $\alpha$ is present 
in the basic definition (2.1) and in the two cases the $r^6$ order 
leading terms of $J^{\bA\bB}_r$ in vacuum are just minus the others; 
or to fix $\alpha$ once and for all for both constructions in the same 
way (e.g. by $\alpha=-1$ as we did in [36,37]). We will see that in 
the latter case $\alpha=1$ would be a slightly more natural choice 
because the {\it anti-holomorphic} rather than the holomorphic 
constructions, both for the energy-momentum and spin-angular momentum, 
fit to the structure of the {\it future} null infinity. However, 
keeping in mind these observations, we retain our previous conventions, 
simply to facilitate the comparison with the previous works. 

In the rest of the paper the 2-surface will be assumed to be a 
large sphere of radius $r$ near the future null infinity in the sense 
defined in section 3 below, and the corresponding spin-angular 
momentum will be denoted by $J^{\bA\bB}_r$. Then, according to the 
general philosophy of the large sphere calculations [38,39], we expand 
the spinor components $\lambda^{\bA}_{\uA}$, the GHP spin coefficients 
and the area element ${\rm d}{\cal S}_r$ as series of ${1\over r}$. 
The expansion coefficients as functions of the remaining coordinates 
depend on the actual construction (holomorphic or anti-holomorphic). 
If therefore $\lambda^{\bA}_{\uA}=:\lambda^{\bA}_{\uA}{}^{(0)}+{1\over 
r}\lambda^{\bA}_{\uA}{}^{(1)}+{1\over r^2}\lambda^{\bA}_{\uA}{}^{(2)}+
...$ and ${\rm d}{\cal S}_r=:r^2(1+{1\over r}s^{(1)}+{1\over r^2}s^{(2)}
+...){\rm d}{\cal S}$, where ${\rm d}{\cal S}$ is the area element on 
the unit sphere ${\cal S}$, then (2.2) takes the form\footnote{*}{We use 
two different notations for the expansion coefficients: $f^{(k)}$ (i.e. 
when the index $k$ is between parentheses) denotes the coefficient of 
$r^{-k}$ in the expansion, which may turn out later to be zero. On the 
other hand, as is usual in the relevant literature, $f\,{}^k$ will 
denote the $(k+1)$th {\it nonvanishing} expansion coefficient of the 
function $f=f({1\over r})$. However, for a function $f$ both $f\,{}^k$ 
and $f^{(k)}$ will never appear.} 

$$\eqalign{
J^{\bA\bB}_r={1\over8\pi G}\oint_{\cal S}&\Bigl\{r^2\Bigl(\lambda
 ^{\bA}_0{}^{(0)}\lambda^{\bB}_1{}^{(0)}+\lambda^{\bA}_1{}^{(0)}\lambda
 ^{\bB}_0{}^{(0)}\Bigl)+r\Bigl(\lambda^{\bA}_0{}^{(0)}\lambda^{\bB}_1{}
 ^{(1)}+\lambda^{\bA}_0{}^{(1)}\lambda^{\bB}_1{}^{(0)}+\lambda^{\bA}_1
 {}^{(0)}\lambda^{\bB}_0{}^{(1)}+\lambda^{\bA}_1{}^{(1)}\lambda^{\bB}_0
 {}^{(0)}\Bigr)+\cr
+&\Bigl(\lambda^{\bA}_0{}^{(0)}\lambda^{\bB}_1{}^{(2)}+\lambda^{\bA}_0
 {}^{(1)}\lambda^{\bB}_1{}^{(1)}+\lambda^{\bA}_0{}^{(2)}\lambda^{\bB}_1
 {}^{(0)}+\lambda^{\bA}_1{}^{(0)}\lambda^{\bB}_0{}^{(2)}+\lambda^{\bA}
 _1{}^{(1)}\lambda^{\bB}_0{}^{(1)}+\lambda^{\bA}_1{}^{(2)}\lambda^{\bB}
 _0{}^{(0)}\Bigr)+...\Bigr\}\times\cr
\times&\Bigl(1+{1\over r}s^{(1)}+{1\over r^2}s^{(2)}+...\Bigr){\rm d}
 {\cal S}.\cr}\eqno(2.5)
$$
\ni
Therefore the $r\rightarrow\infty$ limit of $J^{\bA\bB}_r$ exists iff 
the $r^2$ and $r$ order terms integrate to zero.  Similarly, 

$$\eqalign{
P^{{\bA}{\bB}'}_r={\gamma\over4\pi G}\oint_{\cal S}\Bigl\{
r\Bigl(&\rho'{}^{(1)}\lambda^{\bA}_0{}^{(0)}\bar\lambda^{{\bB}'}_{0'}
  {}^{(0)}+\rho^{(1)}\lambda^{\bA}_1{}^{(0)}\bar\lambda^{{\bB}'}_{1'}
  {}^{(0)}\Bigl)+\cr
+\Bigl(&\rho'{}^{(1)}\bigl(\lambda^{\bA}_0{}^{(0)}\bar\lambda^{{\bB}'}
  _{0'}{}^{(1)}+\lambda^{\bA}_0{}^{(1)}\bar\lambda^{{\bB}'}_{0'}{}
  ^{(0)}\bigr)+\rho'{}^{(2)}\lambda^{\bA}_0{}^{(0)}\bar\lambda^{{\bB}'}
  _{0'}{}^{(0)}+\rho^{(1)}\bigl(\lambda^{\bA}_1{}^{(0)}\bar\lambda
  ^{{\bB}'}_{1'}{}^{(1)}+\cr
&+\lambda^{\bA}_1{}^{(1)}\bar\lambda^{{\bB}'}_{1'}{}^{(0)}\bigr)+\rho
  ^{(2)}\lambda^{\bA}_1{}^{(0)}\bar\lambda^{{\bB}'}_{1'}{}^{(0)}\Bigr)+\cr
+{1\over r}\Bigl(&\rho'{}^{(1)}\bigl(\lambda^{\bA}_0{}^{(0)}\bar\lambda
  ^{{\bB}'}_{0'}{}^{(2)}+\lambda^{\bA}_0{}^{(1)}\bar\lambda^{{\bB}'}
  _{0'}{}^{(1)}+\lambda^{\bA}_0{}^{(2)}\bar\lambda^{{\bB}'}_{0'}{}^{(0)}
  \bigr)+\rho'{}^{(2)}\bigl(\lambda^{\bA}_0{}^{(0)}\bar\lambda^{{\bB}'}
  _{0'}{}^{(1)}+\lambda^{\bA}_0{}^{(1)}\bar\lambda^{{\bB}'}_{0'}{}^{(0)}
  \bigr)+\cr
&+\rho'{}^{(3)}\lambda^{\bA}_0{}^{(0)}\bar\lambda^{{\bB}'}_{0'}{}^{(0)}+
  \rho{}^{(1)}\bigl(\lambda^{\bA}_1{}^{(0)}\bar\lambda^{{\bB}'}_{1'}{}
  ^{(2)}+\lambda^{\bA}_1{}^{(1)}\bar\lambda^{{\bB}'}_{1'}{}^{(1)}+\lambda
  ^{\bA}_1{}^{(2)}\bar\lambda^{{\bB}'}_{1'}{}^{(0)}\bigr)+\cr
&+\rho{}^{(2)}\bigl(\lambda^{\bA}_1{}^{(0)}\bar\lambda^{{\bB}'}_{1'}{}
  ^{(1)}+\lambda^{\bA}_1{}^{(1)}\bar\lambda^{{\bB}'}_{1'}{}^{(0)}\bigr)+
  \rho{}^{(3)}\lambda^{\bA}_1{}^{(0)}\bar\lambda^{{\bB}'}_{1'}{}^{(0)}
  \Bigr)+...\Bigr\}\times\cr
\times\Bigl(1&+{1\over r}s^{(1)}+{1\over r^2}s^{(2)}+...\Bigr){\rm d}
 {\cal S},\cr}\eqno(2.6)
$$
\ni
which has a finite limit iff its $r$ order term integrates to zero. 
\bigskip
\bigskip

\ni
{\lbf 3. The geometry of large spheres}\par
\bigskip
\ni
We assume that the spacetime is future weakly asymptotically simple 
[40], and let $(u,r,\zeta,\bar\zeta)$ be the standard Bondi coordinate 
system in a neighbourhood of the future null infinity ${\cal I}^+$ 
based on a cut ${\cal S}_0$ of ${\cal I}^+$ (the `origin') (see also 
[48,49]). However, following [38], we adapt the null tetrad (and the 
spin frame) to the foliation of the outgoing null hypersurfaces by the 
2-spheres ${\cal S}_{(u,r)}:=\{u={\rm const},\,r={\rm const}\}$, which 
are called the {\it large spheres}. (In particular, the origin ${\cal 
S}_0$ above can be thought of as the $r\rightarrow\infty$ limit of the 
large spheres ${\cal S}_{(0,r)}$.) Explicitly, let $o_A\bar o_{A'}:=l
_a:=\nabla_au$, which is geodesic, and impose the condition $l^a\nabla
_ao_B=0$. Let $n^a=\iota^A\bar\iota^{A'}$ be the other future directed 
null normal to ${\cal S}_{(u,r)}$, normalized by $n^al_a=1$. Then $m^a
=o^A\bar\iota^{A'}$ and $\bar m^a=\iota^A\bar o^{A'}$ are tangent to 
${\cal S}_{(u,r)}$, and the conditions above fix the basis. With this 
choice for the tetrad we have the following restrictions on the GHP 
spin coefficients $\kappa=\varepsilon=\rho-\bar\rho=\tau+\bar\beta'-
\beta=\rho'-\bar\rho'=\tau'-\beta'+\bar\beta=0$. Therefore, $\thorn=
(\partial/\partial r)$, $\edth f=(\delta-(p-q)\beta-q\tau)f$ and 
$\edth' f=(\bar\delta+(p-q)\bar\beta-p\bar\tau)f$ for any scalar $f$ 
of type $(p,q)$, where $\delta:=m^a\nabla_a=:P(\partial/\partial\bar
\zeta)+Q(\partial/\partial\zeta)$, and $\zeta$, $\bar\zeta$ are the 
standard complex stereographic coordinates on the 2-sphere. In 
particular, on the unit sphere in the Minkowski spacetime the edth and 
edth-prime operators take the form ${}_0{\edth}f:={1\over\sqrt2}(1+
\zeta\bar\zeta)(\partial f/\partial\bar\zeta)+{1\over2\sqrt2}(p-q)
\zeta f$ and ${}_0{\edth}'f:={1\over\sqrt2}(1+\zeta\bar\zeta)(\partial 
f/\partial\zeta)-{1\over2\sqrt2}(p-q)\bar\zeta f$, which will be used 
in the subsequent calculations. Overdot will denote partial derivative 
with respect to $u$. 

In the coordinate system $(u,w,\zeta,\bar\zeta)$, where $w:=r^{-1}$ 
and the future null infinity is given by $w=0$, the BMS vector fields 
have the form 
$k^a=(H+(c^{\bi}+\bar c^{\bi})t_{\bi}\,u)({\partial\over\partial u})^a+
c^{\bi}\xi_{\bi}({\partial\over\partial\zeta})^a+\bar c^{\bi}\bar\xi
_{\bi}({\partial\over\partial\bar\zeta})^a+O(r^{-1})$, where $H$ is an 
arbitrary real valued function of $\zeta$ and $\bar\zeta$, $c^{\bi}\in
{\bf C}$, ${\bi}=1,2,3$, and $t_1:=-(\zeta+\bar\zeta)(1+\zeta\bar\zeta)
^{-1}$, $t_2:={\rm i}(\zeta-\bar\zeta)(1+\zeta\bar\zeta)^{-1}$, $t_3:=
(1-\zeta\bar\zeta)(1+\zeta\bar\zeta)^{-1}$ and $\xi_1:=(1-\zeta^2)$, 
$\xi_2:={\rm i}(1+\zeta^2)$ and $\xi_3:=2\zeta$. For the generators 
of the BMS supertranslations $c^{\bi}=0$, and, in particular, the 
independent BMS 
translations take the form $t^a_{\ba}=t_{\ba}({\partial\over\partial 
u})^a+O(r^{-1})$, ${\ba}=0,...,3$, where $t_0:=1$ and $t_{\bi}$ for 
${\bi}=1,2,3$ are given explicitly above. The functions $t_{\ba}$ can 
be written as $t_{\ba}=\sigma^{\bA{\bB}'}_{\ba}\tau_{\bA}\bar\tau
_{{\bB}'}$, where $\tau_0:=\exp({\rm i}\alpha)\root4\of{2}(1+\zeta\bar
\zeta)^{-{1\over2}}$ and $\tau_1:=-\exp({\rm i}\alpha)\root4\of{2}
\zeta(1+\zeta\bar\zeta)^{-{1\over2}}$, and $\exp({\rm i}\alpha)$ is 
an unspecified phase. In Minkowski spacetime the standard constant 
orthonormal frame field $\{E^a_{\ba}\}$ (i.e. the translational Killing 
vectors $K^a_{\ba}$) has precisely this asymptotic form, and if the 
phase $\exp({\rm i}\alpha)$ is chosen to be $-{\rm i}$ then the 
functions $\tau_{\bA}$ are just ${\cal E}^A_{\bA}o_A$, the 
contractions of the Cartesian spin frame with the GHP spin vector 
$o_A$. In fact, in terms of the GHP spin frame $\varepsilon^A_{\uA}=
\{o^A,\iota^A\}$ the Cartesian spin frame ${\cal E}^A_{\bA}=\{O^A,I^A
\}$ has the form 

$$
O^A=-{{\rm i}\over\root4\of{2}}{1\over\sqrt{1+\zeta\bar\zeta}}\Bigl(
  \bar\zeta o^A+\sqrt{2}\iota^A\Bigr), \hskip 12pt
I^A=-{{\rm i}\over\root4\of{2}}{1\over\sqrt{1+\zeta\bar\zeta}}\Bigl(
  o^A-\sqrt{2}\zeta\iota^A\Bigr). \eqno(3.1)
$$
\ni
Thus the BMS translations above are properly normalized. In the 
conformal approach (see, e.g. [40]) the GHP spin frame is rescaled by 
$\hat o^A=\Omega^{-1}o^A$ and $\hat\iota^A=\iota^A$, thus expressing 
$o^A$ and $\iota^A$ by the non-physical spin vectors $\hat o^A$ and 
$\hat\iota^A$ in (3.1), for the asymptotic behaviour of the Cartesian 
spin frame (i.e. the spinor constituents of the translation Killing 
vectors) we directly obtain ${\cal E}^A_{\bA}=\tau_{\bA}\hat\iota^A
+O(\Omega)$, where the $\exp({\rm i}\alpha)=-{\rm i}$ choice was made 
and the conformal factor is $\Omega=r^{-1}$. To find the proper 
normalization of the boost-rotation BMS generators too, let us 
consider the Minkowski spacetime again. The anti-self-dual part 
in the name indices of the boost-rotation Killing vectors $K^a_{\ba
\bb}$ of the Minkowski spacetime is $K^a_{\bA\bB}={1\over\sqrt{2}}
\sigma^{\bi}_{\bA\bB}(K^a_{0{\bi}}+{{\rm i}\over2}\varepsilon_{\bi}{}
^{\bj\bk}K^a_{\bj\bk})$, where ${\bi},{\bj},{\bk}=1,2,3$ and $\sigma
^{\bA}_{\bi}{}_{\bB}:=\sqrt{2}\sigma^{\bA}_{\bi}{}_{{\bB}'}\sigma
^{{\bB}'}_0{}_{\bB}$, the standard $SU(2)$-Pauli matrices, and 
$\varepsilon_{\bi\bj\bk}$ is the alternating Levi-Civita symbol, whose 
indices are moved by the constant (negative definite) metric $\eta
_{\bi\bj}:={\rm diag}\,(-1,-1,-1)$. They yield on ${\cal I}^+$ the BMS 
vector fields 
$k^a_{\bA\bB}:=K^a_{\bA\bB}\vert_{{\cal I}^+}={1\over\sqrt{2}}\sigma
^{\bi}_{\bA\bB}(ut_{\bi}({\partial\over\partial u})^a+\xi_{\bi}({\partial
\over\partial\zeta})^a)+O(r^{-1})=\sigma^{\bi}_{\bA\bB}({1\over\sqrt{2}}
ut_{\bi}\hat\iota^A\bar{\hat\iota}{}^{A'}+\xi_{\bi}(1+\zeta\bar\zeta)
^{-1}\hat\iota^A\bar{\hat o}{}^{A'})+O(\Omega)$. 
Note that the functions $\xi_{\bi}(1+\zeta\bar\zeta)^{-1}$ can 
be written as $\xi_{\bi}(1+\zeta\bar\zeta)^{-1}=-\sigma_{\bi}^{\bA\bB}
\tau_{\bA}\tau_{\bB}$, where the phase in $\tau_{\bA}$ has been (and in 
the rest of the paper will be) chosen as $\exp({\rm i}\alpha)=-{\rm i}$. 
Therefore, apart from the (in general super) translation content, the 
generator of the anti-self-dual rotations at ${\cal I}^+$ is $\sigma^
{\bi}_{\bA\bB}\xi_{\bi}(1+\zeta\bar\zeta)^{-1}=-\tau_{({\bA}}\tau
_{{\bB})}$, i.e. minus the symmetrized product of the functions $\tau
_{\bA}$. In the Appendix we discuss how the generators of the 
translations and of the anti-self-dual rotations can be recovered from 
the solutions of  various spinor equations on the cuts of ${\cal I}^+$. 

       Suppose that, at least in a neighbourhood of ${\cal I}^+$, the 
only matter field that we have is the electromagnetic field, represented 
by the components of the Maxwell spinor (see e.g. [48,49]). Imposing 
slightly stronger fall-off conditions than that coming from the 
definition of the future weak asymptotic simplicity, viz. assuming 
that $\psi_0=r^{-5}\psi_0{}^0+r^{-6}\psi_0{}^1+O(r^{-7})$ and $\phi_0=
r^{-3}\phi_0{}^0+r^{-4}\phi_0{}^1+O(r^{-5})$, Shaw found the asymptotic 
solution of the Einstein--Maxwell equations [38]. We need this solution 
with accuracy $O(r^{-3})$. In particular, 

$$\eqalignno{
P&={1\over\sqrt2}\bigl(1+\zeta\bar\zeta\bigr)\Bigl({1\over r}+{1\over 
  r^3}\sigma^0\bar\sigma^0\Bigr)+O(r^{-5}), &(3.2)\cr
Q&={1\over\sqrt2}\bigl(1+\zeta\bar\zeta\bigr)\Bigl(-{1\over r^2}\sigma
  ^0+{1\over r^4}\bigl({1\over6}\psi_0{}^0-(\sigma^0)^2\bar\sigma^0
  \bigr)\Bigr)+O(r^{-5}). &(3.3)\cr}
$$
\ni
This implies that the area element of the large sphere of radius $r$ 
is ${\rm d}{\cal S}_{(u,r)}=r^2(1-r^{-2}\sigma^0\bar\sigma^0+O(r^{-4}))
{\rm d}{\cal S}$; i.e. in equation (2.4) $s^{(1)}=0$ and $s^{(2)}=-
\sigma^0\bar\sigma^0$. The spin coefficients with definite $(p,q)$ type 
are 

$$\eqalignno{
\sigma=&{1\over r^2}\sigma^0+O(r^{-4}), &(3.4)\cr
\rho=&-{1\over r}-{1\over r^3}\sigma^0\bar\sigma^0+O(r^{-5}), 
   &(3.5)\cr
\sigma'=&-{1\over r}\dot{\bar\sigma^0}-{1\over r^2}\Bigl({1\over2}
   \bar\sigma^0-\,{}_0{\edth}'\,{}_0{\edth}\bar\sigma^0\Bigr)-{1\over 
   r^3}\Bigl(\sigma^0\bar\sigma^0\dot{\bar\sigma^0}+{1\over2}\bar
   \sigma^0\psi_2{}^0+{1\over2}\,{}_0{\edth}'\bar\psi_{1'}{}^0+\cr
&+\bar\sigma^0\bigl(\,{}_0{\edth}^2\bar\sigma^0+\,{}_0{\edth}'{}^2
  \sigma^0\bigr)+\,{}_0{\edth}'\bar\sigma^0\,{}_0{\edth}'\sigma^0-\,
  {}_0{\edth}\bar\sigma^0\,{}_0{\edth}\bar\sigma^0-G\phi_2{}^0\bar
  \phi_{0'}{}^0\Bigr)+O(r^{-4}), &(3.6)\cr
\rho'=&{1\over2r}+{1\over r^2}\Bigl(\psi_2{}^0+\sigma^0\dot{\bar
   \sigma^0}+\,{}_0{\edth}^2\bar\sigma^0\Bigr)-{1\over r^3}\Bigl(\,
   {}_0{\edth}\bar\sigma^0\,{}_0{\edth}'\sigma^0-{1\over2}\sigma^0
   \bar\sigma^0+\sigma^0\,{}_0{\edth}'{}_0{\edth}\bar\sigma^0+\cr
 &+\bar\sigma^0\,{}_0{\edth}\,{}_0{\edth}'\sigma^0+{1\over2}\,{}_0
   {\edth}'\psi_1{}^0+{1\over2}\,{}_0{\edth}\bar\psi_{1'}{}^0-2G\phi
   _1{}^0\bar\phi_{1'}{}^0\Bigr)+O(r^{-4}), &(3.7)\cr
\tau=&{1\over r^2}\,{}_0{\edth}'\sigma^0-{1\over r^3}\Bigl(2\sigma^0\,
   {}_0{\edth}\bar\sigma^0+\psi_1{}^0\Bigr)+O(r^{-4}); &(3.8)\cr}
$$
\ni
while the spin coefficient representing the GHP connection 1-form is 

$$
\beta=-{1\over r}{1\over2\sqrt2}\zeta-{1\over r^2}{1\over2\sqrt2}\bar
\zeta\sigma^0-{1\over r^3}\Bigl({1\over2\sqrt2}\zeta\sigma^0\bar\sigma^0
+{1\over2}\psi_1{}^0+\sigma^0\,{}_0{\edth}\bar\sigma^0\Bigr)+O(r^{-4}).
\eqno(3.9)
$$
\ni
Finally, for the Weyl and Maxwell spinor components one has the familiar 
peeling: $\psi_n=r^{n-5}\psi_n{}^0+O(r^{n-6})$, $n=0,...,4$, and $\phi
_n=r^{n-3}\phi_n{}^0+O(r^{n-4})$, $n=0,1,2$, where 
$\psi_4{}^0=-\ddot{\bar\sigma^0}$, 
$\psi_3{}^0=-\,{}_0{\edth}\dot{\bar\sigma^0}$ and 
$\psi_2{}^0-\bar\psi_{2'}{}^0=\bar\sigma^0\dot{\sigma^0}-\sigma^0\dot{
\bar\sigma^0}+\,{}_0{\edth}'{}^2\sigma^0-\,{}_0{\edth}^2\bar\sigma^0$. 

      The spacetime will be called stationary, if it admits a timelike 
Killing vector field, at least on an open neighbourhood of ${\cal I}^+$, 
which can be extended to ${\cal I}^+$ into a BMS translation [48,38]. 
Then there is a cut, chosen to be the new origin, whose asymptotic 
shear is vanishing, and the asymptotic shear on the cut given by the 
supertranslation $S:{\cal S}\rightarrow{\bf R}$ with respect to the 
new origin is $\sigma^0=-\,{}_0{\edth}^2S$. Then, after an appropriate 
translation of the new origin, the asymptotic value of the Weyl and 
Maxwell spinor components that we need in what follows on such a cut 
take the form $\psi_1{}^0=-3G\,{}_0{\edth}(MS+{\rm i}J)$, $\psi_2{}^0=
-GM$ and $\phi_1{}^0={1\over2}(e+{\rm i}\mu)$. Here $M$, $e$ and $\mu$ 
are real constants, interpreted as the total mass, the total electric 
charge and the total magnetic charge, respectively, and $J$ is a real 
function with structure $J=\sum_{m=-1}^{m=1}J_mY_{1,m}$ for some 
constants $J_0$ and $J_{\pm1}$, where $Y_{1,m}$ are the standard $j=1$ 
spherical harmonics. Rewriting $J$ by the familiar polar coordinates 
$(\theta,\phi)$ (defined by $\zeta=:\cot{\theta\over2}\exp({\rm i}
\phi)$) into the form $J=j_1\sin\theta\cos\phi+j_2\sin\theta\sin\phi+
j_3\cos\theta$, the real constants $j_{\bi}=(j_1,j_2,j_3)$ defined in 
this way are interpreted as the components of the total spatial angular 
momentum of the stationary solution. In particular, for the Kerr--Newman 
solution $j_1=j_2=0$ and $j_3=Ma$, and hence on the shear-free $u={\rm 
const}.$ cuts $\psi_1{}^0=-{3\over\sqrt{2}}{\rm i}\,GMa\sin\theta\exp(
{\rm i}\phi)$. (Its apparent deviation from the expression given in 
[17,38] by the factor $-\exp({\rm i}\phi)$ is a consequence of the 
different choices for the holomorphic coordinates on ${\cal S}$: Our 
choice is $\bar\zeta$ [even if spherical polar coordinates are present], 
while in [17,38] it is $-\log\bar\zeta=\log\tan{\theta\over2}+{\rm i}
\phi$.)
\bigskip
\bigskip

\ni
{\lbf 4. The anti-holomorphic spin-angular momentum}\par
\bigskip
\ni
Substituting the expansion $\lambda_{\uA}=:\lambda_{\uA}{}^{(0)}+{1
\over r}\lambda_{\uA}{}^{(1)}+{1\over r^2}\lambda_{\uA}{}^{(2)}+...$ 
of the spinor components and the expressions (3.2-9) for the 
functions $P$ and $Q$ and the spin coefficients into the equations 
defining the anti-holomorphic spinor fields on ${\cal S}_{(u,r)}$, we 
obtain the following hierarchy of equations 

$$\eqalignno{
{}_0\edth\lambda_1{}^{(0)}&+{1\over2}\lambda_0{}^{(0)}=0,&(4.1.a)\cr
{}_0\edth\lambda_0{}^{(0)}&=0,&(4.1.b)\cr
{}_0\edth\lambda_1{}^{(1)}&+{1\over2}\lambda_0{}^{(1)}=-\bigl(\psi_2
   {}^0+\sigma^0\dot{\bar\sigma{}^0}+{}_0\edth^2\bar\sigma^0\bigr)
   \lambda_0{}^{(0)},&(4.2.a)\cr
{}_0\edth\lambda_0{}^{(1)}&=0,&(4.2.b)\cr
{}_0\edth\lambda_1{}^{(2)}&+{1\over2}\lambda_0{}^{(2)}=\sigma^0\,{}_0
   {\edth}'\lambda_1{}^{(1)}-\bigl(\psi_2{}^0+\sigma^0\dot{\bar\sigma
   {}^0}+{}_0\edth^2\bar\sigma^0\bigr)\lambda_0{}^{(1)}+\bigl(\sigma^0
   \,{}_0\edth\bar\sigma^0+{1\over2}\psi_1{}^0\bigr)\lambda_1{}^{(0)}+\cr
+\bigl(&\,{}_0{\edth}\bar\sigma^0\,{}_0{\edth}'\sigma^0+\sigma^0\,{}_0
   {\edth}'{}_0{\edth}\bar\sigma^0+\bar\sigma^0\,{}_0{\edth}\,{}_0
   {\edth}'\sigma^0+{1\over2}\,{}_0{\edth}'\psi_1{}^0+{1\over2}\,{}_0
   {\edth}\bar\psi_{1'}{}^0-2G\phi_1{}^0\bar\phi_{1'}{}^0\bigr)\lambda
   _0{}^{(0)},&(4.3.a)\cr
{}_0\edth\lambda_0{}^{(2)}&=\sigma^0\bigl({}_0\edth'\lambda_0{}^{(1)}
   -\lambda_1{}^{(1)}\bigr)-\bigl(\sigma^0\,{}_0\edth\bar\sigma^0+{1
   \over2}\psi_1{}^0\bigr)\lambda_0{}^{(0)}. &(4.3.b)\cr}
$$
\ni
Thus by (4.1.a-b) the zeroth order spinor components are just the 
components of the constant spinors of Minkowski spacetime, and hence, 
in addition to (4.1.a-b), they satisfy ${}_0\edth'\lambda_1{}^{(0)}=
0$ and ${}_0\edth'\lambda_0{}^{(0)}=\lambda_1{}^{(0)}$, too. Therefore, 
there are precisely two solutions $\lambda_{\uA}^{\bA}{}^{(0)}$, ${\bA}
=0,1$, of (4.1.a-b), and we choose them to be given explicitly by 
(A.2.2) with $\rho'={1\over2r}$ (see the Appendix). 

Since the left hand side of (4.1), (4..2) and (4.3) are the same, the 
solution in each order will be the sum of a particular solution and 
the general zeroth order (i.e. the constant) solution. Therefore, the 
general $r^{-2}$ accurate anti-holomorphic spinor fields form a six 
rather than the expected two complex dimensional space. In fact, in the 
small sphere calculations [37] we had similar spurious solutions, due 
to the fact that there is no natural isomorphism between the space of 
the anti-holomorphic spinor fields on two different two-surfaces. Thus 
they represent a gauge ambiguity in the first and second order 
corrections, and the physical quantities must be invariant with respect 
to the substitutions $\lambda_{\uA}{}^{(1)}\mapsto\lambda_{\uA}{}^{(1)}
+C_{\bA}\lambda_{\uA}^{\bA}{}^{(0)}$ and $\lambda_{\uA}{}^{(2)}\mapsto
\lambda_{\uA}{}^{(2)}+D_{\bA}\lambda_{\uA}^{\bA}{}^{(0)}$ for any 
constants $C_{\bA}$ and $D_{\bA}$. 

To compute the spin-angular momentum based on equation (2.5), first 
observe that the integral of the $r^2$ order term is vanishing, because 
that is just the spin-angular momentum in Minkowski spacetime. However, 
the $r$ order term of the integrand is 

$$\eqalign{
&\lambda^{\bA}_0{}^{(0)}\lambda^{\bB}_1{}^{(1)}+\lambda^{\bA}_0{}^{(1)}
 \lambda^{\bB}_1{}^{(0)}+\lambda^{\bA}_1{}^{(0)}\lambda^{\bB}_0{}^{(1)}
 +\lambda^{\bA}_1{}^{(1)}\lambda^{\bB}_0{}^{(0)}=\cr
&=-2\,{}_0{\edth}\bigl(\lambda_1^{\bA}{}^{(0)}\lambda_1^{\bB}{}^{(1)}+
 \lambda_1^{\bB}{}^{(0)}\lambda_1^{\bA}{}^{(1)}\bigr)+2\lambda_1^{\bA}
 {}^{(0)}\bigl(\,{}_0{\edth}\lambda_1^{\bB}{}^{(1)}+{1\over2}\lambda_0
 ^{\bB}{}^{(1)}\bigr)+2\lambda_1^{\bB}{}^{(0)}\bigl(\,{}_0{\edth}
 \lambda_1^{\bA}{}^{(1)}+{1\over2}\lambda_0^{\bA}{}^{(1)}\bigr)=\cr
&=-2\,{}_0{\edth}\bigl(\lambda_1^{\bA}{}^{(0)}\lambda_1^{\bB}{}^{(1)}+
 \lambda_1^{\bB}{}^{(0)}\lambda_1^{\bA}{}^{(1)}\bigr)-2\Bigl(\psi_2{}^0
 +\sigma^0\dot{\bar\sigma^0}+\,{}_0{\edth}^2\bar\sigma^0\Bigr)\bigl(
 \lambda_1^{\bA}{}^{(0)}\lambda_0^{\bB}{}^{(0)}+\lambda_1^{\bB}{}^{(0)}
 \lambda_0^{\bA}{}^{(0)}\bigr),\cr}
$$
\ni
where we used equations (4.1) and (4.2); furthermore, by (4.1) the last 
of the second term is a total divergence: 

$$
{}_0{\edth}^2\bar\sigma^0\bigl(\lambda_1^{\bA}{}^{(0)}\lambda_0^{\bB}
{}^{(0)}+\lambda_1^{\bB}{}^{(0)}\lambda_0^{\bA}{}^{(0)}\bigr)=\,{}_0
{\edth}\Bigl(\,{}_0{\edth}\bar\sigma^0\bigl(\lambda_1^{\bA}{}^{(0)}
\lambda_0^{\bB}{}^{(0)}+\lambda_1^{\bB}{}^{(0)}\lambda_0^{\bA}{}^{(0)}
\bigr)+\bar\sigma^0\lambda_0^{\bA}{}^{(0)}\lambda_0^{\bB}{}^{(0)}\Bigr).
\eqno(4.4)
$$
\ni
Thus for the anti-holomorphic spinor fields $J^{\bA\bB}_r$ is diverging 
at ${\cal I}^+$ unless the integral 

$$
L^{\bA\bB}:=-{1\over4\pi G}\oint_{\cal S}\bigl(\psi_2{}^0+\sigma
^0\dot{\bar\sigma^0}\bigr)\Bigl(\lambda_1^{\bA}{}^{(0)}\lambda_0^{\bB}
{}^{(0)}+\lambda_0^{\bA}{}^{(0)}\lambda_1^{\bB}{}^{(0)}\Bigr){\rm d}
{\cal S} \eqno(4.5)
$$
\ni
is vanishing. (We return to the discussion of $L^{\bA\bB}$ below.) Next 
let us consider the $r^0$ order term of (2.5):  
       
$$\eqalignno{
{1\over2}\oint_{\cal S}\Bigl(&\lambda^{\bA}_0{}^{(0)}\lambda^{\bB}_1{}
 ^{(2)}+\lambda^{\bA}_0{}^{(1)}\lambda^{\bB}_1{}^{(1)}+\lambda^{\bA}_0
 {}^{(2)}\lambda^{\bB}_1{}^{(0)}+\lambda^{\bA}_1{}^{(0)}\lambda^{\bB}
 _0{}^{(2)}+\lambda^{\bA}_1{}^{(1)}\lambda^{\bB}_0{}^{(1)}+\lambda
 ^{\bA}_1{}^{(2)}\lambda^{\bB}_0{}^{(0)}-\cr
&-\sigma^0\bar\sigma^0\bigl(\lambda^{\bA}_0{}^{(0)}\lambda_1^{\bB}{}
 ^{(0)}+\lambda^{\bA}_1{}^{(0)}\lambda_0^{\bB}{}^{(0)}\bigl)\Bigr){\rm 
 d}{\cal S}=\cr 
=\oint_{\cal S}\Bigl(&-\bigl(\,{}_0{\edth}\lambda_1^{\bA}{}^{(0)}\bigr)
 \lambda_1^{\bB}{}^{(2)}+\lambda_1^{\bB}{}^{(1)}\bigl(-\,{}_0{\edth}
 \lambda_1^{\bA}{}^{(1)}-(\psi_0{}^0+\sigma^0\dot{\bar\sigma^0}+\,{}_0
 {\edth}^2\bar\sigma^0)\lambda_0^{\bA}{}^{(0)}\bigr)+{1\over2}\lambda
 _0^{\bA}{}^{(2)}\lambda_1^{\bB}{}^{(0)}-\cr
&-\bigl(\,{}_0{\edth}\lambda_1^{\bB}{}^{(0)}\bigr)\lambda_1^{\bA}{}
 ^{(2)}+\lambda_1^{\bA}{}^{(1)}\bigl(-\,{}_0{\edth}\lambda_1^{\bB}{}
 ^{(1)}-(\psi_0{}^0+\sigma^0\dot{\bar\sigma^0}+\,{}_0{\edth}^2\bar
 \sigma^0)\lambda_0^{\bB}{}^{(0)}\bigr)+{1\over2}\lambda_0^{\bB}{}
 ^{(2)}\lambda_1^{\bA}{}^{(0)}-\cr
&-{1\over2}\sigma^0\bar\sigma^0\bigl(\lambda^{\bA}_0{}^{(0)}\lambda_1
 ^{\bB}{}^{(0)}+\lambda^{\bA}_1{}^{(0)}\lambda_0^{\bB}{}^{(0)}\bigl)
 \Bigr){\rm  d}{\cal S}=\cr 
=\oint_{\cal S}\Bigl(&\lambda_1^{\bA}{}^{(0)}\bigl(\,{}_0{\edth}\lambda
 _1^{\bB}{}^{(2)}+{1\over2}\lambda_0^{\bB}{}^{(2)}-{1\over2}\sigma^0\bar
 \sigma^0\lambda_0^{\bB}{}^{(0)}\bigr)+\lambda_1^{\bB}{}^{(0)}\bigl(\,
 {}_0{\edth}\lambda_1^{\bA}{}^{(2)}+{1\over2}\lambda_0^{\bA}{}^{(2)}-
 {1\over2}\sigma^0\bar\sigma^0\lambda_0^{\bA}{}^{(0)}\bigr)-\cr
&-\bigl(\psi_0{}^0+\sigma^0\dot{\bar\sigma^0}+\,{}_0{\edth}^2\bar\sigma
 ^0\bigr)\bigl(\lambda_0^{\bA}{}^{(0)}\lambda_1^{\bB}{}^{(1)}+\lambda_1
 ^{\bA}{}^{(1)}\lambda_0^{\bB}{}^{(0)}\bigr)\Bigr){\rm d}{\cal S}=\cr
=\oint_{\cal S}\Bigl(&-\rho'{}^{(3)}\bigl(\lambda_0^{\bA}{}^{(0)}\lambda
 _1^{\bB}{}^{(0)}+\lambda_1^{\bA}{}^{(0)}\lambda_0^{\bB}{}^{(0)}\bigr)+\cr
&+\bigl(\psi_1{}^0+2\sigma^0\,{}_0{\edth}\bar\sigma^0\bigr)\lambda_1
 ^{\bA}{}^{(0)}\lambda_1^{\bB}{}^{(0)}+\sigma^0\bigl(\lambda_1^{\bA}{}
 ^{(0)}\,{}_0{\edth}'\lambda_1^{\bB}{}^{(1)}+\lambda_1^{\bB}{}^{(0)}\,
 {}_0{\edth}'\lambda_1^{\bA}{}^{(1)}\bigr)-\cr
&-\bigl(\psi_2{}^0+\sigma^0\dot{\bar\sigma^0}+\,{}_0{\edth}^2\bar\sigma
 ^0\bigr)\bigl(\lambda_0^{\bA}{}^{(0)}\lambda_1^{\bB}{}^{(1)}+\lambda_0
 ^{\bA}{}^{(1)}\lambda_1^{\bB}{}^{(0)}+\lambda_1^{\bA}{}^{(0)}\lambda_0
 ^{\bB}{}^{(1)}+\lambda_1^{\bA}{}^{(1)}\lambda_0^{\bB}{}^{(0)}\bigr)
\Bigr){\rm d}{\cal S}, &(4.6)\cr}
$$
\ni
where $\rho'{}^{(3)}$ is the 3rd order term of $\rho'$ in (3.7), and 
we used (4.1.a), (4.2.a) and (4.3.a). Substituting (3.7) here, using 
(4.1.b) and the consequences ${}_0{\edth}'\lambda_1^{\bA}{}^{(0)}=0$ 
and ${}_0{\edth}'\lambda_0^{\bA}{}^{(0)}=\lambda_1^{\bA}{}^{(0)}$ of 
(4.1), the integral of the first three terms of the right hand side 
of (4.6) can be written as 

$$\eqalignno{
\oint_{\cal S}\Bigl(&\bigl(\psi_1{}^0+2\sigma^0\,{}_0{\edth}\bar\sigma
 ^0\bigr)\lambda_1^{\bA}{}^{(0)}\lambda_1^{\bB}{}^{(0)}-\,{}_0{\edth}'
 \sigma^0\bigl(\lambda_1^{\bA}{}^{(0)}\lambda_1^{\bB}{}^{(1)}+\lambda
 _1^{\bB}{}^{(0)}\lambda_1^{\bA}{}^{(1)}\bigr)+\bigl(\,{}_0{\edth}\bar
 \sigma^0\,{}_0{\edth}'\sigma^0-{1\over2}\sigma^0\bar\sigma^0+\cr
&+\sigma^0\,{}_0{\edth}'{}_0{\edth}\bar\sigma^0+\bar\sigma^0\,{}_0
 {\edth}\,{}_0{\edth}'\sigma^0+{1\over2}\,{}_0{\edth}'\psi_1{}^0+{1
 \over2}\,{}_0{\edth}\bar\psi_{1'}{}^0-2G\phi_1{}^0\bar\phi_{1'}{}^0
 \bigr)\bigl(\lambda_0^{\bA}{}^{(0)}\lambda_1^{\bB}{}^{(0)}+\lambda_1
 ^{\bA}{}^{(0)}\lambda_0^{\bB}{}^{(0)}\bigr)\Bigr){\rm d}{\cal S}=\cr
=\oint_{\cal S}\Bigl(&\bigl({1\over2}\bar\psi_{1'}{}^0+\bar\sigma^0
 \,{}_0{\edth}'\sigma^0\bigr)\lambda_0^{\bA}{}^{(0)}\lambda_0^{\bB}
 {}^{(0)}-\,{}_0{\edth}'\sigma^0\bigl(\lambda_1^{\bA}{}^{(0)}\lambda
 _1^{\bB}{}^{(1)}+\lambda_1^{\bB}{}^{(0)}\lambda_1^{\bA}{}^{(1)}
 \bigr)-\cr
&-\bigl(\,{}_0{\edth}'\sigma^0\,{}_0{\edth}\bar\sigma^0+{1\over2}
 \sigma^0\bar\sigma^0+2G\phi_1{}^0\bar\phi_{1'}{}^0\bigr)\bigl(\lambda
 _0^{\bA}{}^{(0)}\lambda_1^{\bB}{}^{(0)}+\lambda_1^{\bA}{}^{(0)}
 \lambda_0^{\bB}{}^{(0)}\bigr)\Bigr){\rm d}{\cal S}=\cr
={1\over2}\oint_{\cal S}\Bigl(&\bigl(\bar\psi_{1'}{}^0+2\bar\sigma^0
 \,{}_0{\edth}'\sigma^0+\,{}_0{\edth}'(\sigma^0\bar\sigma^0)\bigr)
 \lambda_0^{\bA}{}^{(0)}\lambda_0^{\bB} {}^{(0)}-4G\,{}_0{\edth}'
 (\phi_1{}^0\bar\phi_{1'}{}^0)\bigl(\lambda_0^{\bA}{}^{(0)}\lambda_1
 ^{\bB}{}^{(0)}+\lambda_1^{\bA}{}^{(0)}\lambda_0^{\bB}{}^{(0)}\bigr)-\cr
&-2\,{}_0{\edth}'\sigma^0\bigl(\lambda_1^{\bA}{}^{(0)}(\lambda_1^{\bB}
 {}^{(1)}+\lambda^{\bB}_0{}^{(0)}\,{}_0{\edth}\bar\sigma^0)+\lambda_1
 ^{\bB}{}^{(0)}(\lambda_1^{\bA}{}^{(1)}+\lambda^{\bA}_0{}^{(0)}\,
 {\edth}\bar\sigma^0)\bigr)\Bigr){\rm d}{\cal S}.&(4.7)\cr}
$$
\ni
Using (4.2.a) again, the integral of the last term on the right hand 
side of (4.6) can be rewritten as 

$$\eqalignno{
-\oint_{\cal S}&\bigl(\psi_2{}^0+\sigma^0\dot{\bar\sigma^0}+\,{}_0
 {\edth}^2\bar\sigma^0\bigr)\Bigl(\lambda_0^{\bA}{}^{(0)}\lambda_1
 ^{\bB}{}^{(1)}+\lambda_0^{\bA}{}^{(1)}\lambda_1^{\bB}{}^{(0)}+\lambda
 _1^{\bA}{}^{(0)}\lambda_0^{\bB}{}^{(1)}+\lambda_1^{\bA}{}^{(1)}
 \lambda_0^{\bB}{}^{(0)}\Bigr){\rm d}{\cal S}=\cr
=\oint_{\cal S}\Bigl(&\lambda_1^{\bB}{}^{(1)}\bigl(\,{}_0{\edth}
 \lambda_1^{\bA}{}^{(1)}+{1\over2}\lambda_0^{\bA}{}^{(1)}\bigr)+
 \lambda_1^{\bA}{}^{(1)}\bigl(\,{}_0{\edth}\lambda_1^{\bB}{}^{(1)}+
 {1\over2}\lambda_0^{\bB}{}^{(1)}\bigr)-\cr
&-\bigl(\psi_2{}^0+\sigma^0\dot{\bar\sigma^0}+\,{}_0{\edth}^2\bar
 \sigma^0\bigr)\bigl(\lambda_0^{\bA}{}^{(1)}\lambda_1^{\bB}{}^{(0)}+
 \lambda_1^{\bA}{}^{(0)}\lambda_0^{\bB}{}^{(1)}\bigr)\Bigr){\rm d}
 {\cal S}=\cr
=\oint_{\cal S}\Bigl\{&\lambda_0^{\bA}{}^{(1)}\Bigl({1\over2}\lambda
 _1^{\bB}{}^{(1)}-\bigl(\psi_2{}^0+\sigma^0\dot{\bar\sigma^0}+\,{}_0
 {\edth}^2\bar\sigma^0\bigr)\lambda_1^{\bB}{}^{(0)}\Bigr)+\cr
&+\lambda_0^{\bB}{}^{(1)}\Bigl({1\over2}\lambda_1^{\bA}{}^{(1)}-\bigl(
 \psi_2{}^0+\sigma^0\dot{\bar\sigma^0}+\,{}_0{\edth}^2\bar\sigma^0
\bigr)\lambda_1^{\bA}{}^{(0)}\Bigr)\Bigr\}{\rm d}{\cal S}.&(4.8)\cr}
$$
\ni
Therefore, $\lambda_0^{\bA}{}^{(2)}$ and $\lambda_1^{\bA}{}^{(2)}$ are 
not needed to calculate the anti-holomorphic spin-angular momentum, 
but $\lambda_0^{\bA}{}^{(1)}$ and $\lambda_1^{\bA}{}^{(1)}$ do appear 
in (4.6) explicitly. Since, however, the physical quantities should 
not be sensitive to the addition of the spurious zeroth order 
solutions to $\lambda_0^{\bA}{}^{(1)}$ and $\lambda_1^{\bA}{}^{(1)}$, 
and by such solutions $\lambda_0^{\bA}{}^{(1)}=0$ can always be achieved, 
whenever (4.8) gives zero, that should be a gauge term. To see that, 
in fact, this is the case, recall that (4.1.b) and (4.2.b) are 
the same, thus we may write $\lambda_0^{\bA}{}^{(1)}=C^{\bA}{}_{\bC}
\lambda_0^{\bC}{}^{(0)}=-2C^{\bA}{}_{\bC}\,{}_0{\edth}\lambda_1^{\bC}
{}^{(0)}$ for some constant complex $2\times2$ matrix $C^{\bA}{}
_{\bC}$. (However, $C^{\bA}{}_{\bB}$ is not quite arbitrary, that is 
restricted by the requirement that the pair of anti-holomorphic spinor 
fields should form a normalized spin frame. In fact, from $\varepsilon
^{\bA\bB}=\varepsilon^{\uA\uB}(\lambda^{\bA}_{\uA}{}^{(0)}+{1\over r}
\lambda^{\bA}_{\uA}{}^{(1)}+...)(\lambda^{\bB}_{\uB}{}^{(0)}+{1\over r}
\lambda^{\bB}_{\uB}{}^{(1)}+...)$ it follows that $C^{\bA\bB}$ must 
be symmetric.) Substituting this into (4.8) and using (4.2.a) we obtain

$$\eqalignno{
-2\bigl(C^{\bA}{}_{\bC}\delta^{\bB}_{\bD}+C^{\bB}{}_{\bC}\delta^{\bA}
 _{\bD}\bigr)\oint_{\cal S}&\,{}_0{\edth}\lambda_1^{\bC}{}^{(0)}\Bigl(
 {1\over2}\lambda_1^{\bD}{}^{(1)}-\bigl(\psi_2{}^0+\sigma^0\dot{\bar
 \sigma^0}+\,{}_0{\edth}^2\bar\sigma^0\bigr)\lambda_1^{\bD}{}^{(0)}
 \Bigr){\rm d}{\cal S}=\cr
=\bigl(C^{\bA}{}_{\bC}\delta^{\bB}_{\bD}+C^{\bB}{}_{\bC}\delta^{\bA}
 _{\bD}\bigr)\oint_{\cal S}&\lambda_1^{\bC}{}^{(0)}\Bigl(\,{}_0{\edth}
 \lambda_1^{\bD}{}^{(1)}-2\,{}_0{\edth}\bigl(\psi_2{}^0+\sigma^0\dot
 {\bar\sigma^0}+\,{}_0{\edth}^2\bar\sigma^0\bigr)\lambda_1^{\bD}{}^{(0)}
 +\cr
&+\bigl(\psi_2{}^0+\sigma^0\dot{\bar\sigma^0}+\,{}_0{\edth}^2\bar\sigma
 ^0\bigr)\lambda_0^{\bD}{}^{(0)}\Bigr){\rm d}{\cal S}=\cr
=\bigl(C^{\bA}{}_{\bC}\delta^{\bB}_{\bD}+C^{\bB}{}_{\bC}\delta^{\bA}
 _{\bD}\bigr)\oint_{\cal S}&\lambda_1^{\bC}{}^{(0)}\Bigl(-{1\over2}
 \lambda_0^{\bD}{}^{(1)}-2\,{}_0{\edth}\bigl(\psi_2{}^0+\sigma^0\dot
 {\bar\sigma^0}+\,{}_0{\edth}^2\bar\sigma^0\bigr)\lambda_1^{\bD}{}^{(
 0)}\Bigr){\rm d}{\cal S}=\cr
={1\over4}\bigl(C^{\bA}{}_{\bC}C^{\bB}{}_{\bD}+C^{\bB}{}_{\bC}C^{\bA}&
 {}_{\bD}\bigr)\oint_{\cal S}\lambda_0^{\bC}{}^{(0)}\lambda_1^{\bD}{}
 ^{(0)}{\rm d}{\cal S}+\cr
-\bigl(C^{\bA}{}_{\bC}\delta^{\bB}_{\bD}+C^{\bB}{}_{\bC}\delta^{\bA}
 _{\bD}\bigr)\oint_{\cal S}&\bigl(\psi_2{}^0+\sigma^0\dot{\bar\sigma^0}
 +\,{}_0{\edth}^2\bar\sigma^0\bigr)\Bigl(\lambda_0{}^{\bC}{}^{(0)}
 \lambda_1{}^{\bD}{}^{(0)}+\lambda_1{}^{\bC}{}^{(0)}\lambda_0{}^{\bD}
 {}^{(0)}\Bigr){\rm d}{\cal S}.&(4.9)\cr}
$$
\ni
However, the first integral on the right hand side of (4.9) is 
vanishing, while, taking into account (4.4), the second is seen 
to be proportional to $L^{\bA\bB}$ above. Therefore, the 
anti-holomorphic spin-angular momentum for the large sphere of radius 
$r$ is 

$$
J^{\bA\bB}_r=rL^{{\bA}{\bB}}+\bigl(C^{\bA}{}_{\bC}\delta^{\bB}_{\bD}+
C^{\bB}{}_{\bC}\delta^{\bA}_{\bD}\bigr)L^{\bC\bD}-E^{\bA\bB}+I^{\bA\bB}
+O(r^{-1}),\eqno(4.10)
$$
\ni
where we introduced the notations 

$$\eqalignno{
E^{\bA\bB}:=&{1\over2\pi}\oint_{\cal S}\phi_1{}^0\bar\phi_{1'}{}^0
 \Bigl(\lambda^{\bA}_0{}^{(0)}\lambda^{\bB}_1{}^{(0)}+\lambda^{\bA}
 _1{}^{(0)}\lambda^{\bB}_1{}^{(0)}\Bigr){\rm d}{\cal S}, &(4.11) \cr
I^{\bA\bB}:=&{1\over8\pi G}\oint_{\cal S}\Bigl\{\Bigl(\bar\psi_{1'}
 {}^{0}+2\bar\sigma^0\,{}_0{\edth}'\sigma^0+\,{}_0{\edth}'\bigl(
 \sigma^0\bar\sigma^0\bigr)\Bigr)\lambda_0^{\bA}{}^{(0)}\lambda_0
 ^{\bB}{}^{(0)}-\cr
&-2\,{}_0{\edth}'\sigma^0\Bigl(\lambda_1^{\bA}{}^{(0)}\bigl(\lambda_1
 ^{\bB}{}^{(1)}+\lambda^{\bB}_0{}^{(0)}\,{}_0{\edth}\bar\sigma^0\bigr)+
 \lambda_1^{\bB}{}^{(0)}\bigl(\lambda_1^{\bA}{}^{(1)}+\lambda^{\bA}_0
 {}^{(0)}\,{}_0{\edth}\bar\sigma^0\bigr)\Bigr)\Bigr\}{\rm d}{\cal S}. 
 &(4.12)\cr}
$$
\ni
Therefore, as we noted above, the quasi-local spin-angular momentum 
based on Bramson's superpotential and the anti-holomorphic spinors is 
diverging at the future null infinity, furthermore, its finite part 
is ambiguous unless $L^{\bA\bB}$ is vanishing. In addition, contrary 
to expectations, the electromagnetic field contributes to $J^{\bA\bB}
_r$ in the $O(1)$ order. On the other hand, the first three terms 
together in the integrand of $I^{\bA\bB}$ is just the integrand of 
Bramson's specific spin-angular momentum expression based on the 
asymptotic twistor equation [22]. Though in the present case the 
spinor fields $\lambda^{\bA}_A$ are anti-holomorphic and not the 
solutions of the asymptotic twistor equations, the zeroth order part 
of $\lambda^{\bA}_A$, which appears as the coefficient of the first 
three terms of the integrand of $I^{\bA\bB}$, coincides with the 
zeroth order part of the solutions of the asymptotic twistor 
equations. Furthermore, the coefficient $\lambda^{\bA}_0{}^{(0)}
\lambda^{\bB}_0{}^{(0)}$ is just minus the generator of the 
anti-self-dual rotations on the $u=0$ cut (the `origin') of ${\cal I}
^+$ (see Section 3. and the Appendix). It might be worth emphasizing 
that these generators of the anti-self-dual rotation BMS vector fields 
emerged naturally, like the approximate rotation-boost Killing vectors 
in the small sphere calculations [37], without putting them into the 
general formulae by hand. Although $I^{\bA\bB}$ depends on the 
solution $\lambda_1^{\bA}{}^{(1)}$ of (4.2.a) (and hence its integrand 
is a genuinely non-local expression), that is independent of the gauge 
solutions. In fact, using ${}_0{\edth}'\lambda_1^{\bA}{}^{(0)}=0$, it 
is easy to see that the addition of a gauge solution to $\lambda_1^{\bA}
{}^{(1)}$ changes the integrand by a total ${}_0{\edth}'$-derivative. 
To clarify the meaning of the term of the integrand involving $\lambda
_1^{\bA}{}^{(1)}$, recall that the BMS vector fields $k^a_{\bA\bB}$ are 
tangent only to the origin cut, and they can be represented completely 
by $\lambda_0^{\bA}{}^{(0)}\lambda_0^{\bB}{}^{(0)}$ only there. On 
general cuts $k^a_{\bA\bB}$ contain correction terms proportional to 
the generator $({\partial\over\partial u})^a$ of ${\cal I}^+$. Then the 
integral of the term involving $\lambda_1^{\bA}{}^{(1)}$ can be 
considered as the contribution of this correction term built in a 
non-local way from the fields on the actual cut and the spinor 
constituents of the BMS rotations on the origin cut.

However, $L^{\bA\bB}$ is not zero in general, because its components 
are proportional to that of the spatial part of the Bondi--Sachs 
energy-momentum 

$$
{}_\infty P^{\bA{\bB}'}:=-{1\over4\pi G}\oint_{\cal S}\bigl(\psi_2{}^0
+\sigma^0\dot{\bar\sigma^0}\bigr)\lambda_0^{\bA}{}^{(0)}\bar\lambda
_{0'}^{{\bB}'}{}^{(0)}{\rm d}{\cal S}. \eqno(4.13)
$$
\ni
To see this, let us substitute the explicit solutions (A.2.2) into 
(4.5). We obtain $L^{\bf 00}=-\sqrt{2}\,{}_\infty P^{{\bf 0}{\bf 
1}'}$, $L^{\bf 01}={1\over\sqrt{2}}(\,{}_\infty P^{{\bf 0}{\bf 0}'}-
\,{}_\infty P^{{\bf 1}{\bf 1}'})$ and $L^{\bf 11}=\sqrt{2}\,{}_\infty 
P^{{\bf 1}{\bf 0}'}$, i.e. $L^{\bA\bB}$ represents the linear momentum. 
In fact, for ${\bA}={\bB}=0$, ${\bA}={\bB}=1$ and ${\bA}=0$, ${\bB}=1$ 
the coefficient of $\psi_2{}^0+\sigma^0\dot{\bar\sigma^0}$ in (4.5) is 
$-2\zeta(1+\zeta\bar\zeta)^{-1}$, $2\bar\zeta(1+\zeta\bar\zeta)^{-1}$ 
and $(\zeta\bar\zeta-1)(1+\zeta\bar\zeta)^{-1}$, respectively, which 
are proportional to the independent spatial BMS translations. Therefore, 
the anti-holomorphic spin-angular momentum can be finite only in the 
center-of-mass system (i.e. when the spatial components of the 
Bondi--Sachs energy-momentum are vanishing), and hence $I^{\bA\bB}$ in 
the $O(1)$ part of (4.10) appears to represent only the intrinsic or 
{\it spin part} of the total angular momentum, while $rL^{\bA\bB}$ 
appears to be the orbital part of the angular momentum. 
To check whether this interpretation is correct we should calculate 
the quasi-local Pauli--Lubanski spin (2.4) built from the quasi-local 
anti-holomorphic Dougan--Mason energy-momentum $P^{\bA{\bB}'}$ and the 
anti-holomorphic spin-angular momentum. However, to compute the spin, 
we need to know the Dougan--Mason energy-momentum for large spheres 
with $O(r^{-1})$ accuracy. Since this has been calculated only in 
stationary spacetimes [39] (where a physical term was apparently 
overlooked and the gauge ambiguity caused by the spurious solutions 
was not considered at all), first we must calculate this.

      The $r$ order part of (2.6) is vanishing, because $\oint_{\cal S}
\lambda^{\bA}_0{}^{(0)}\bar\lambda^{{\bB}'}_{0'}{}^{(0)}{\rm d}{\cal S}=
2\oint_{\cal S}\lambda^{\bA}_1{}^{(0)}\bar\lambda^{{\bB}'}_{1'}{}^{(0)}
{\rm d}{\cal S}=4\pi\,\sigma^{{\bA}{\bB}'}_0$ (see Appendix A.2), 
i.e. $P^{{\bA}{\bB}'}_r$ has a finite $r\rightarrow\infty$ limit at 
${\cal I}^+$. Substituting (3.5) and (3.7) into the finite term of 
(2.6), using (4.1.a), ${}_0{\edth}'\lambda^{\bA}_0{}^{(0)}=\lambda^{\bA}
_1{}^{(0)}$ and its complex conjugate, (4.2.a) and the fact that $\psi_2
{}^0+\sigma^0\dot{\bar\sigma^0}+\,{}_0{\edth}^2\bar\sigma^0$ is real, 
we obtain 

$$\eqalign{
\oint_{\cal S}\Bigl\{{1\over2}&\Bigl(\lambda^{\bA}_0{}^{(0)}\bar\lambda
 ^{{\bB}'}_{0'}{}^{(1)}+\lambda^{\bA}_0{}^{(1)}\bar\lambda^{{\bB}'}_{0'}
 {}^{(0)}\Bigr)-\Bigl(\lambda^{\bA}_1{}^{(0)}\bar\lambda^{{\bB}'}_{1'}
 {}^{(1)}+\lambda^{\bA}_1{}^{(1)}\bar\lambda^{{\bB}'}_{1'}{}^{(0)}\Bigr)
 +\cr
+&\Bigl(\psi_2{}^0+\sigma^0\dot{\bar\sigma^0}+\,{}_0{\edth}^2\bar\sigma
 ^0\Bigr)\lambda^{\bA}_0{}^{(0)}\bar\lambda^{{\bB}'}_{0'}{}^{(0)}\Bigr\}
 {\rm d}{\cal S}=\cr
=\oint_{\cal S}\Bigl\{&\bar\lambda^{{\bB}'}_{0'}{}^{(0)}\Bigl(\,{}_0
 {\edth}\lambda^{\bA}_1{}^{(1)}+{1\over2}\lambda^{\bA}_0{}^{(1)}\Bigr)+
 \lambda^{\bA}_0{}^{(0)}\Bigl(\,{}_0{\edth}'\bar\lambda^{{\bB}'}_{1'}{}
 ^{(1)}+{1\over2}\bar\lambda^{{\bB}'}_{0'}{}^{(1)}\Bigr)+\cr
+&\Bigl(\psi_2{}^0+\sigma^0\dot{\bar\sigma^0}+\,{}_0{\edth}^2\bar\sigma
 ^0\Bigr)\lambda^{\bA}_0{}^{(0)}\bar\lambda^{{\bB}'}_{0'}{}^{(0)}\Bigr\}
 {\rm d}{\cal S}=\cr
=-\oint_{\cal S}&\Bigl(\psi_2{}^0+\sigma^0\dot{\bar\sigma^0}+\,{}_0
 {\edth}^2\bar\sigma^0\Bigr)\lambda^{\bA}_0{}^{(0)}\bar\lambda^{{\bB}'}
 _{0'}{}^{(0)}{\rm d}{\cal S}.\cr}
$$
\ni
Since the last term of the integrand is a total ${}_0{\edth}$-derivative: 
$({}_0{\edth}^2\bar\sigma^0)\lambda^{\bA}_0{}^{(0)}\bar\lambda^{{\bB}'}
_{0'}{}^{(0)}={}_0{\edth}(\,{}_0{\edth}\bar\sigma^0\,\lambda^{\bA}_0{}
^{(0)}\bar\lambda^{{\bB}'}_{0'}{}^{(0)})-({}_0{\edth}\bar\sigma^0)
\lambda^{\bA}_0{}^{(0)}\,{}_0{\edth}\bar\lambda^{{\bB}'}_{0'}{}^{(0)}=
{}_0{\edth}(\,{}_0{\edth}\bar\sigma^0\,\lambda^{\bA}_0{}^{(0)}\bar
\lambda^{{\bB}'}_{0'}{}^{(0)}-\bar\sigma^0\lambda^{\bA}_0{}^{(0)}\bar
\lambda^{{\bB}'}_{1'}{}^{(0)})$, the finite part of (2.6) is, in fact, 
the Bondi--Sachs energy-momentum (4.13).

The $O(r^{-1})$ term of (2.6) is 

$$\eqalign{
\oint_{\cal S}\Bigl(&\bigl(\rho'{}^{(3)}-{1\over2}\sigma^0\bar\sigma^0
  \bigr)\lambda^{\bA}_0{}^{(0)}\bar\lambda^{{\bB}'}_{0'}{}^{(0)}+\bigl(
  \psi_2{}^0+\sigma^0\dot{\bar\sigma^0}+\,{}_0{\edth}^2\bar\sigma^0\bigr)
  \bigl(\lambda^{\bA}_0{}^{(0)}\bar\lambda^{{\bB}'}_{0'}{}^{(1)}+\lambda
  ^{\bA}_0{}^{(1)}\bar\lambda^{{\bB}'}_{0'}{}^{(0)}\bigr)+\cr
+{1\over2}&\bigl(\lambda^{\bA}_0{}^{(0)}\bar\lambda^{{\bB}'}_{0'}{}^{(2)}
  +\lambda^{\bA}_0{}^{(1)}\bar\lambda^{{\bB}'}_{0'}{}^{(1)}+\lambda^{\bA}
  _0{}^{(2)}\bar\lambda^{{\bB}'}_{0'}{}^{(0)}\bigr)-\,{}_0{\edth}'
  \lambda^{\bA}_0{}^{(0)}\bar\lambda^{{\bB}'}_{1'}{}^{(2)}-\lambda^{\bA}
  _1{}^{(2)}\,{}_0{\edth}\bar\lambda^{{\bB}'}_{0'}{}^{(0)}-\lambda^{\bA}
  _1{}^{(1)}\bar\lambda^{{\bB}'}_{1'}{}^{(1)}\Bigr){\rm d}{\cal S}=\cr
=\oint_{\cal S}\Bigl(&\bigl(\rho'{}^{(3)}-{1\over2}\sigma^0\bar\sigma^0
  \bigr)\lambda^{\bA}_0{}^{(0)}\bar\lambda^{{\bB}'}_{0'}{}^{(0)}+\bigl(
  \psi_2{}^0+\sigma^0\dot{\bar\sigma^0}+\,{}_0{\edth}^2\bar\sigma^0\bigr)
  \bigl(\lambda^{\bA}_0{}^{(0)}\bar\lambda^{{\bB}'}_{0'}{}^{(1)}+\lambda
  ^{\bA}_0{}^{(1)}\bar\lambda^{{\bB}'}_{0'}{}^{(0)}\bigr)+\cr
+&\lambda^{\bA}_0{}^{(0)}\bigl({}_0{\edth}'\bar\lambda^{{\bB}'}_{1'}{}
  ^{(2)}+{1\over2}\bar\lambda^{{\bB}'}_{0'}{}^{(2)}\bigr)+\bar\lambda
  ^{{\bB}'}_{0'}{}^{(0)}\bigl({}_0{\edth}\lambda^{\bA}_1{}^{(2)}+{1
  \over2}\lambda^{\bA}_0{}^{(2)}\bigr)+{1\over2}\lambda^{\bA}_0{}^{(1)}
  \bar\lambda^{{\bB}'}_{0'}{}^{(1)}-\lambda^{\bA}_1{}^{(1)}\bar\lambda
  ^{{\bB}'}_{1'}{}^{(1)}\Bigr){\rm d}{\cal S}=\cr
=\oint_{\cal S}\Bigl(-&\bigl(2G\phi_1{}^0\bar\phi_{1'}{}^0+\,{}_0{\edth}
  \bar\sigma^0\,{}_0{\edth}'\sigma^0\bigr)\lambda^{\bA}_0{}^{(0)}\bar
  \lambda^{{\bB}'}_{0'}{}^{(0)}-\cr
-&\bigl(\,{}_0{\edth}\bar\sigma^0\bigr)\lambda^{\bA}_0{}^{(0)}\bar\lambda
  ^{{\bB}'}_{1'}{}^{(1)}-\bigl(\,{}_0{\edth}'\sigma^0\bigr)\lambda^{\bA}
  _1{}^{(1)}\bar\lambda^{{\bB}'}_{0'}{}^{(0)}+{1\over2}\lambda^{\bA}_0{}
  ^{(1)}\bar\lambda^{{\bB}'}_{0'}{}^{(1)}-\lambda^{\bA}_1{}^{(1)}\bar
  \lambda^{{\bB}'}_{1'}{}^{(1)}\Bigr){\rm d}{\cal S}=\cr
=\oint_{\cal S}\Bigl(-&2G\phi_1{}^0\bar\phi_{1'}{}^0\lambda^{\bA}_0{}
  ^{(0)}\bar\lambda^{{\bB}'}_{0'}{}^{(0)}+{1\over2}\lambda^{\bA}_0{}
  ^{(1)}\bar\lambda^{{\bB}'}_{0'}{}^{(1)}-\bigl(\lambda^{\bA}_1{}^{(1)}+
  \lambda^{\bA}_0{}^{(0)}\,{}_0{\edth}\bar\sigma^0\bigr)\bigl(\bar\lambda
  ^{{\bB}'}_{1'}{}^{(1)}+\bar\lambda^{{\bB}'}_{0'}{}^{(0)}\,{}_0{\edth}'
  \sigma^0\bigr)\Bigr){\rm d}{\cal S}.\cr}\eqno(4.14)
$$
\ni
Here first we used ${}_0{\edth}'\lambda^{\bA}_0{}^{(0)}=\lambda^{\bA}_1
{}^{(0)}$, and then (4.3.a), (4.1.b), (3.7) and the fact that $\psi_2{}
^0+\sigma^0\dot{\bar\sigma^0}+\,{}_0{\edth}^2\bar\sigma^0$ is real. 
However, as we saw above, $\lambda^{\bA}_0{}^{(1)}=C^{\bA}{}_{\bC}
\lambda^{\bC}_0{}^{(0)}=-2C^{\bA}{}_{\bC}\,{}_0{\edth}\lambda^{\bC}_1{}
^{(0)} $, and by means of which (4.2.a) takes the form ${}_0{\edth}
(\lambda^{\bA}_1{}^{(1)}+\lambda^{\bA}_0{}^{(0)}\,{}_0{\edth}\bar\sigma
^0)=C^{\bA}{}_{\bC}\,{}_0{\edth}\lambda^{\bC}_1{}^{(0)}-(\psi_2{}^0+
\sigma^0\dot{\bar\sigma^0})\lambda^{\bA}_0{}^{(0)}$. Since ${}_0{\edth}$ 
acting on $s=-{1\over2}$ spin weight quantities is isomorphism (see e.g. 
[47,50]), we can write $\lambda^{\bA}_1{}^{(1)}+\lambda^{\bA}_0{}^{(0)}
\,{}_0{\edth}\bar\sigma^0=C^{\bA}{}_{\bC}\lambda^{\bC}_1{}^{(0)}-\,{}_0
{\edth}^{-1}((\psi_2{}^0+\sigma^0\dot{\bar\sigma^0})\lambda^{\bA}_0{}
^{(0)})$. Substituting these into (4.12), using ${}_0{\edth}'\lambda
^{\bA}_0{}^{(0)}=\lambda^{\bA}_1{}^{(0)}$, (4.13) and the expression for 
the $L_2$-scalar product of the components $\lambda^{\bA}_0{}^{(0)}$ and 
$\lambda^{\bA}_1{}^{(0)}$ given in Appendix A.2, finally we obtain 

$$
P^{{\bA}{\bB}'}_r={}_\infty P^{{\bA}{\bB}'}+{1\over r}\bigl(C^{\bA}{}
_{\bC}\delta^{{\bB}'}_{{\bD}'}+\bar C^{{\bB}'}{}_{{\bD}'}\delta^{\bA}
_{\bC}\bigr){}_\infty P^{{\bC}{\bD}'}-{1\over r}F^{{\bA}{\bB}'}+
{1\over r}M^{{\bA}{\bB}'}+O({1\over r^2}), \eqno(4.15)
$$
\ni
where we used the notations 

$$\eqalignno{
F^{{\bA}{\bB}'}:=&{1\over2\pi}\oint_{\cal S}\phi_1{}^0\bar\phi_{1'}
 {}^0\lambda^{\bA}_0{}^{(0)}\bar\lambda^{{\bB}'}_{0'}{}^{(0)}{\rm d}
 {\cal S}, &(4.16)\cr
M^{{\bA}{\bB}'}:=&-{1\over4\pi G}\oint_{\cal S}\,{}_0{\edth}^{-1}
 \Bigl(\bigl(\psi_2{}^0+\sigma^0\dot{\bar\sigma^0}\bigr)\lambda^{\bA}
 _0{}^{(0)}\Bigr)\,{}_0{\edth}'^{-1}\Bigl(\bigl(\bar\psi_{2'}{}^0+
 \bar\sigma^0\dot{\sigma^0}\bigr)\bar\lambda^{{\bB}'}_{0'}{}^{(0)}
 \Bigr){\rm d}{\cal S}.&(4.16)\cr}
$$
\ni
Therefore, as it could be expected, the $r^{-1}$ order term in the 
expansion of the anti-holomorphic Dougan--Mason energy-momentum is 
ambiguous, and, in addition to the electromagnetic contribution, 
there is an extra term. Note that $E^{\bA\bB}$ above is related to 
$F^{{\bA}{\bB}'}$ like $L^{\bA\bB}$ to ${}_\infty P^{{\bA}{\bB}'}$, 
and hence $E^{\bf 00}=-\sqrt{2}F^{{\bf 0}{\bf 1}'}$, $E^{\bf 01}={1
\over\sqrt{2}}(F^{{\bf 0}{\bf 0}'}-F^{{\bf 1}{\bf 1}'})$ and $E^{\bf 
11}=\sqrt{2}F^{{\bf 1}{\bf 0}'}$.

Substituting (4.10) and (4.15) into (2.4) and using how the components 
of $L^{\bA\bB}$ and  $E^{\bA\bB}$ are related to those of ${}_\infty 
P^{{\bA}{\bB}'}$ and $F^{{\bA}{\bB}'}$, respectively, we obtain 

$$
S_{{\bA}{\bB}'}={\rm i}\Bigl({}_\infty P_{{\bA}{\bC}'}\bar I^{{\bC}'}
{}_{{\bB}'}+M_{{\bA}{\bC}'}\bar L^{{\bC}'}{}_{{\bB}'}-{}_\infty P
_{{\bB}'{\bC}}I^{\bC}{}_{\bA}-M_{{\bB}'{\bC}}L^{\bC}{}_{\bA}\Bigr).
\eqno(4.17)
$$
\ni
The diverging term, the ambiguities and the contribution of the 
electromagnetic field disappeared. Therefore, {\it the quasi-local 
Pauli--Lubanski spin vector built from the anti-holomorphic 
Dougan--Mason energy-momentum and the anti-holomorphic spin-angular 
momentum (2.2) has finite limit at the future null infinity, it is 
free of ambiguities, and is built only from the gravitational data}. 
Note that our Pauli--Lubanski spin is free of the so-called 
supertranslation ambiguity, because this is defined in terms of the 
solutions of an elliptic differential equation on the cut in question, 
and not by means of the BMS boost-rotation vector fields. Thus the 
present construction is similar in its spirit to that of Penrose [12]. 
The fact that we could derive only a Pauli--Lubanski spin is compatible 
with the idea of Bergmann and Thomson [21] that the gravitational 
angular momentum should be analogous to spin (justifying the 
`spin-angular momentum' terminology), but raises the question as 
whether that should be completed by an orbital angular momentum part 
or not. Another interesting issue is how the Pauli--Lubanski spin 
changes under (infinitesimal) supertranslations, in particular, under 
time translations; or whether there exists a flux for $S_{{\bA}{\bA}'}$ 
through ${\cal I}^+$ or not. However, these questions are beyond the 
scope of the present paper. 
\bigskip
\bigskip

\ni
{\lbf 5. Stationary spacetimes}\par
\bigskip
\ni
Suppose that the spacetime is stationary. First we show that all the 
terms of the integrand of (4.12) involving the asymptotic shear 
together integrates to zero. Bramson already showed that $2\bar
\sigma^0\,{}_0{\edth}'\sigma^0+\,{}_0{\edth}'(\sigma^0\bar\sigma^0)=
{}_0{\edth}'{}^3(S\,{}_0{\edth}^2S+{1\over2}({}_0{\edth}S)^2)-\,{}_0
{\edth}(S\,{}_0{\edth}'{}^3\,{}_0{\edth}S+3\,{}_0{\edth}'{}^2\,{}_0
{\edth}S\,{}_0{\edth}S+3({}_0{\edth}'S)^2)$, which, together with 
${}_0{\edth}\lambda^{\bA}_0{}^{(0)}=0$, ${}_0{\edth}'\lambda^{\bA}_0
{}^{(0)}=\lambda^{\bA}_1{}^{(0)}$ and ${}_0{\edth}'\lambda^{\bA}_1{}
^{(0)}=0$, gives that the first two $\sigma^0$-terms of the integrand 
of (4.12) give zero. 
To evaluate the last term of the integrand, first we must solve 
(4.2.a). In stationary spacetimes (4.2.a) takes the form ${}_0{\edth}
\lambda^{\bA}_1{}^{(1)}+{1\over2}C^{\bA}{}_{\bB}\lambda^{\bB}_0{}^{(0)}
=(GM+\,{}_0{\edth}^2\,{}_0{\edth}'{}^2S)\lambda^{\bA}_0{}^{(0)}$. 
Using ${\rm dim}\,{\rm ker}\,{\edth}_{(-1,0)}=0$, this can be solved 
explicitly. Its solution is $\lambda^{\bA}_1{}^{(1)}={}_0{\edth}(\,{}_0
{\edth}'{}^2S)\lambda^{\bA}_0{}^{(0)}+(-2GM\delta^{\bA}_{\bB}+C^{\bA}{}
_{\bB})\lambda^{\bB}_1{}^{(0)}$. Substituting this into the last term 
of the integrand, that takes the form 

$$\eqalign{
&-2\,{}_0{\edth}'\sigma^0\Bigl(\lambda^{\bA}_1{}^{(0)}\bigl(\lambda
 ^{\bB}_1{}^{(1)}+\lambda^{\bB}_0{}^{(0)}\,{}_0{\edth}\bar\sigma^0\bigr)
 +\lambda^{\bB}_1{}^{(0)}\bigl(\lambda^{\bA}_1{}^{(1)}+\lambda^{\bA}_0
 {}^{(0)}\,{}_0{\edth}\bar\sigma^0\bigr)\Bigr)=\cr
=&-2\Bigl(\bigl(-2GM\delta^{\bA}_{\bC}+C^{\bA}{}_{\bC}\bigr)\delta^{\bB}
 _{\bD}+\bigl(-2GM\delta^{\bB}_{\bD}+C^{\bB}{}_{\bD}\bigr)\delta^{\bA}
 _{\bC}\Bigr)\,{}_0{\edth}'\bigl(\sigma^0\lambda^{\bC}_1{}^{(0)}\lambda
 ^{\bD}_1{}^{(0)}\bigr).\cr}
$$
\ni
Therefore, the last term of the integrand of (4.12) integrates to zero, 
too. By $\psi_2{}^0=-Gm$ the integrand of $L^{\bA\bB}$ is the total ${}
_0{\edth}'$-derivative ${}_0{\edth}'(-Gm\lambda^{\bA}_0{}^{(0)}\lambda
^{\bB}_0{}^{(0)})$, and by $\phi_1{}^{0}={1\over2}(e+{\rm i}\mu)$ the 
integrand of $E^{\bA\bB}$ is ${}_0{\edth}'({1\over4}[e^2+\mu^2]\lambda
^{\bA}_0{}^{(0)}\lambda^{\bB}_0{}^{(0)})$. Thus in stationary 
spacetimes (4.10) reduces to 

$$
J^{\bA\bB}_r={1\over8\pi G}\oint_{\cal S}\bar\psi_{1'}{}^{0}\lambda
^{\bA}_0{}^{(0)}\lambda^{\bB}_0{}^{(0)}{\rm d}{\cal S}+O(r^{-1}), 
\eqno(5.1)
$$
\ni
which is the `standard' expression for the angular momentum in stationary 
spacetimes. In fact, substituting $\psi_1{}^0=-3G\,{}_0{\edth}(MS+{\rm i}
\sum_{m=-1}^{m=1}J_mY_{1,m})$ here and using ${}_0{\edth}'\lambda^{\bA}_0
{}^{(0)}=\lambda^{\bA}_1{}^{(0)}$, we find 

$$
J^{\bA\bB}_r={3M\over8\pi}\oint_{\cal S}S\Bigl(\lambda^{\bA}_0{}^{(0)}
\lambda^{\bB}_1{}^{(0)}+\lambda^{\bA}_1{}^{(0)}\lambda^{\bB}_0{}^{(0)}
\Bigr){\rm d}{\cal S}-{3{\rm i}\over8\pi}\sum^1_{m=-1}\overline{J_m}
\oint_{\cal S}\overline{Y_{1,m}}\Bigl(\lambda^{\bA}_0{}^{(0)}\lambda
^{\bB}_1{}^{(0)}+\lambda^{\bA}_1{}^{(0)}\lambda^{\bB}_0{}^{(0)}\Bigr)
{\rm d}{\cal S}+O(r^{-1}).
$$
\ni
Substituting the specific solutions (A.2.2) here and defining $c_m:=
M\oint_{\cal S}\overline{Y_{1,m}}S{\rm d}{\cal S}$ and the corresponding 
real components $c^{\bi}=(c^1,c^2,c^3)$ by $\sum^{m=1}_{m=-1}c_mY_{1,m}
=:c^1\sin\theta\cos\phi+c^2\sin\theta\sin\phi+c^3\cos\theta$, for the 
leading terms we finally obtain $J^{\bi\bj}=\varepsilon^{\bi\bj\bk}j
_{\bk}$ and $J^{{\bf 0}{\bk}}=-c^{\bk}$. Thus $J^{\bA\bB}$ reproduces 
the $j_{\bk}$'s, and the $c^{\bk}$'s can be interpreted as the 
components of the relativistic center-of-mass. In particular, for the 
Kerr--Newman solution on the shear-free cuts $J^{\bf 12}=-Ma$. 
Using the expressions of Appendix A.2 for the $L_2$ scalar product of 
the spinor components $\lambda^{\bA}_{\uA}{}^{(0)}$, we obtain ${}
_\infty P^{{\bA}{\bB}'}=M\,\sigma^{{\bA}{\bB}'}_0$, $F^{{\bA}{\bB}'}=
{1\over2}(e^2+\mu^2)\,\sigma^{{\bA}{\bB}'}_0$ and $M^{{\bA}{\bB}'}=-2
GM^2\,\sigma^{{\bA}{\bB}'}_0$. Hence in stationary spacetimes the 
Pauli--Lubanski spin reduces to that of Bramson. As he showed [23], 
this is invariant with respect to supertranslations. 
\bigskip
\bigskip

\ni
{\lbf 6. The holomorphic spin-angular momentum}\par
\bigskip
\ni
For the components $\lambda_{\uA}=:\lambda_{\uA}{}^{(0)}+{1\over r}
\lambda_{\uA}{}^{(1)}+{1\over r^2}\lambda_{\uA}{}^{(2)}+...$ of the 
holomorphic spinor fields on ${\cal S}_{(u,r)}$ we obtain the equations 

$$\eqalignno{
{}_0{\edth}'\lambda_0{}^{(0)}&-\lambda_1{}^{(0)}=0,&(6.1.a)\cr
{}_0{\edth}'\lambda_1{}^{(0)}&-\dot{\bar\sigma^0}\lambda_0{}^{(0)}=0,
   &(6.1.b)\cr
{}_0{\edth}'\lambda_0{}^{(1)}&-\lambda_1{}^{(1)}=\,{}_0{\edth}\bigl(
   \bar\sigma^0\lambda_0{}^{(0)}\bigr),&(6.2.a)\cr
{}_0{\edth}'\lambda_1{}^{(1)}&-\dot{\bar\sigma^0}\lambda_0{}^{(1)}=\bar
   \sigma^0\Bigl(\,{}_0{\edth}\lambda_1{}^{(0)}+{1\over2}\lambda_0{}
   ^{(0)}\Bigr)-\,{}_0{\edth}'\Bigl(\bigl(\,{}_0{\edth}\bar\sigma^0
   \bigr)\lambda_0{}^{(0)}\Bigr),&(6.2.b)\cr
{}_0{\edth}'\lambda_0{}^{(2)}&-\lambda_1{}^{(2)}={}_0{\edth}\bigl(
   \bar\sigma^0\lambda_0{}^{(1)}\bigr)-\bigl({1\over2}\bar\psi_{1'}
   {}^0+\bar\sigma^0\,{}_0{\edth}'\sigma^0\bigr)\lambda_0{}^{(0)},
   &(6.3.a)\cr
{}_0{\edth}'\lambda_1{}^{(2)}&-\dot{\bar\sigma^0}\lambda_0{}^{(2)}=
   \bar\sigma^0\Bigl(\,{}_0{\edth}\lambda_1{}^{(1)}+{1\over2}\lambda
   _0{}^{(1)}\Bigr)-\,{}_0{\edth}'\Bigl(\bigl({}_0{\edth}\bar\sigma^0
   \bigr)\lambda_0{}^{(1)}\Bigr)+\Bigl({1\over2}\bar\psi_{1'}{}^0+\bar
   \sigma^0\,{}_0{\edth}'\sigma^0\Bigr)\lambda_1{}^{(0)}+\cr
+&\bar\sigma^0\,{}_0{\edth}\bar\sigma^0\,{}_0{\edth}\lambda_0{}^{(0)}
   +\Bigl({1\over2}\bar\sigma^0\psi_2{}^0+{1\over2}\,{}_0{\edth}'\bar
   \psi_{1'}{}^0+\,{}_0{\edth}'\bigl(\bar\sigma^0\,{}_0{\edth}\sigma^0
   \bigr)+\bar\sigma^0\,{}_0{\edth}^2\bar\sigma^0-G\phi_2{}^0\bar\phi
   _{0'}{}^0\Bigr)\lambda_0{}^{(0)}. &(6.3.b)\cr}
$$
\ni
Therefore, the zeroth order holomorphic spinors are {\it not} constant 
unless $\dot\sigma^0$ is vanishing, and hence the coefficient of the 
$r^2$ order term of (2.5) is not zero in general. Therefore, {\it the 
quasi-local spin-angular momentum (2.2) based the holomorphic spinor 
fields is diverging near ${\cal I}^+$ in presence of outgoing 
gravitational radiation}. 
Thus let us concentrate on spacetimes for which $\dot\sigma^0=0$. Then 
$\lambda_0{}^{(0)}$, $\lambda_1{}^{(0)}$ are components of a constant 
spinor field on ${\cal S}$, implying the vanishing of the $r^2$ order 
term in (2.5), and the vanishing of the first term on the right hand 
side of (6.2.b). Then, however, its general solution is $\lambda^{\bA}
_1{}^{(1)}=-({}_0{\edth}\bar\sigma^0)\lambda^{\bA}_0{}^{(0)}+C^{\bA}{}
_{\bB}\lambda^{\bB}_1{}^{(0)}$ for some $2\times2$ complex matrix $C
^{\bA}{}_{\bB}$, by means of which the solution of (6.2.a) is $\lambda
^{\bA}_0{}^{(1)}=C^{\bA}{}_{\bB}\lambda^{\bB}_0{}^{(0)}$. Therefore, 
the integrand of the $r$ order term of (2.5) is 

$$\eqalign{
&\lambda^{\bA}_0{}^{(0)}\lambda^{\bB}_1{}^{(1)}+\lambda^{\bA}_0{}^{(1)}
 \lambda^{\bB}_1{}^{(0)}+\lambda^{\bA}_1{}^{(0)}\lambda^{\bB}_0{}^{(1)}
 +\lambda^{\bA}_1{}^{(1)}\lambda^{\bB}_0{}^{(0)}=\cr
&=\bigl(\delta^{\bA}_{\bC}C^{\bB}{}_{\bD}+C^{\bA}{}_{\bC}\delta^{\bB}
 _{\bD}\bigr)\bigl(\lambda_0^{\bC}{}^{(0)}\lambda_1^{\bD}{}^{(0)}+
 \lambda_0^{\bD}{}^{(0)}\lambda_1^{\bC}{}^{(0)}\bigr)-2\,{}_0{\edth}
 \bigl(\bar\sigma^0\lambda_0^{\bA}{}^{(0)}\lambda_0^{\bB}{}^{(0)}\bigr).
\cr}
$$
\ni
However, its integral is zero, i.e. {\it in absence of outgoing 
gravitational radiation (i.e. if $\dot\sigma^0=0$) the quasi-local 
angular momentum based on (2.2) and the holomorphic spinor fields has 
a finite limit at the future null infinity}. 

To calculate this finite value let us consider the $r^0$ order term of 
(2.5). Using the explicit solutions for $\lambda^{\bA}_0{}^{(1)}$ and 
$\lambda^{\bA}_1{}^{(1)}$ above, the fact that $\lambda^{\bA}_A{}
^{(0)}$ is constant and (6.3.a) we obtain 

$$\eqalignno{
\oint_{\cal S}\Bigl(&\lambda^{\bA}_0{}^{(0)}\lambda^{\bB}_1{}
 ^{(2)}+\lambda^{\bA}_0{}^{(1)}\lambda^{\bB}_1{}^{(1)}+\lambda^{\bA}_0
 {}^{(2)}\lambda^{\bB}_1{}^{(0)}+\lambda^{\bA}_1{}^{(0)}\lambda^{\bB}
 _0{}^{(2)}+\lambda^{\bA}_1{}^{(1)}\lambda^{\bB}_0{}^{(1)}+\lambda
 ^{\bA}_1{}^{(2)}\lambda^{\bB}_0{}^{(0)}-\cr
&-\sigma^0\bar\sigma^0\bigl(\lambda^{\bA}_0{}^{(0)}\lambda_1^{\bB}{}
 ^{(0)}+\lambda^{\bA}_1{}^{(0)}\lambda_0^{\bB}{}^{(0)}\bigl)\Bigr){\rm 
 d}{\cal S}=\cr 
=\oint_{\cal S}\Bigl(&\bigl(\,{}_0{\edth}'\lambda_0^{\bB}{}^{(0)}\bigr)
 \lambda_0^{\bA}{}^{(2)}+\bigl(\,{}_0{\edth}'\lambda_0^{\bA}{}^{(0)}
 \bigr)\lambda_0^{\bA}{}^{(2)}+\cr
&+C^{\bA}{}_{\bC}\lambda_0^{\bC}{}^{(0)}\bigl(-\,{}_0{\edth}\bar\sigma^0
 \lambda_0^{\bB}{}^{(0)}+C^{\bB}{}_{\bD}\lambda^{\bD}_1{}^{(0)}\bigr)+
 C^{\bB}{}_{\bD}\lambda_0^{\bD}{}^{(0)}\bigl(-\,{}_0{\edth}\bar\sigma
 ^0\lambda_0^{\bA}{}^{(0)}+C^{\bA}{}_{\bC}\lambda^{\bC}_1{}^{(0)}\bigr)
 +\cr
&+\lambda_0^{\bA}{}^{(0)}\lambda_1^{\bB}{}^{(2)}+\lambda_0^{\bB}{}^{(0)}
 \lambda_1^{\bA}{}^{(2)}-\sigma^0\bar\sigma^0\bigl(\lambda_0^{\bA}{}
 ^{(0)}\,{}_0{\edth}'\lambda_0^{\bB}{}^{(0)}+\lambda_0^{\bB}{}^{(0)}\,
 {}_0{\edth}'\lambda_0^{\bA}{}^{(0)}\bigr)\Bigr){\rm d}{\cal S}=\cr
=\oint_{\cal S}\Bigl(&\lambda_0^{\bA}{}^{(0)}\bigl(-\,{}_0{\edth}'
 \lambda_0^{\bB}{}^{(2)}+\lambda_1^{\bB}{}^{(2)}\bigr)+\lambda_0^{\bB}
 {}^{(0)}\bigl(-\,{}_0{\edth}'\lambda_0^{\bA}{}^{(0)}+\lambda_1^{\bA}
 {}^{(2)}\bigr)-\sigma^0\bar\sigma^0\,{}_0{\edth}\bigl(\lambda_0^{\bA}
 {}^{(0)}\lambda_0^{\bB}{}^{(0)}\bigr)-\cr
&-\bigl(C^{\bA}{}_{\bC}\delta^{\bB}_{\bD}+C^{\bB}{}_{\bD}\delta^{\bA}
 _{\bC}\bigr)\,{}_0{\edth}\bar\sigma^0\lambda_0^{\bC}{}^{(0)}\lambda_0
 ^{\bD}{}^{(0)}+C^{\bA}{}_{\bC}C^{\bB}{}_{\bD}\bigl(\lambda^{\bC}_0{}
 ^{(0)}\lambda^{\bD}_1{}^{(0)}+\lambda^{\bC}_1{}^{(0)}\lambda^{\bD}_0
 {}^{(0)}\bigr)\Bigr){\rm d}{\cal S}=\cr
=\oint_{\cal S}\Bigl(&\lambda_0^{\bA}{}^{(0)}\bigl(-\,{}_0{\edth}'
 \lambda_0^{\bB}{}^{(2)}+\lambda_1^{\bB}{}^{(2)}\bigr)+\lambda_0^{\bB}
 {}^{(0)}\bigl(-\,{}_0{\edth}'\lambda_0^{\bA}{}^{(2)}+\lambda_1^{\bA}
 {}^{(2)}\bigr)+\,{}_0{\edth}'\bigl(\sigma^0\bar\sigma^0\bigr)\lambda
 _0^{\bA}{}^{(0)}\lambda_0^{\bB}{}^{(0)}\Bigr){\rm d}{\cal S}=\cr
=\oint_{\cal S}\Bigl(&\bar\psi_{1'}{}^0+2\bar\sigma^0\,{}_0{\edth}'
 \sigma^0+\,{}_0{\edth}'\bigl(\sigma^0\bar\sigma^0\bigr)\Bigr)\lambda
 ^{\bA}_0{}^{(0)}\lambda^{\bB}_0{}^{(0)}{\rm d}{\cal S}, &(6.4)\cr}
$$
\ni
which is precisely Bramson's specific spin-angular momentum expression 
based on the asymptotic twistor equation. Thus, substituting $\sigma^0=-
{}_0{\edth}^2S$ here, finally we obtain (5.1), i.e. {\it for stationary 
spacetimes the holomorphic and the anti-holomorphic constructions give 
the same `standard' expression}. We expect that at the {\it past} null 
infinity the holomorphic construction works properly, and the 
anti-holomorphic diverges in presence of incoming gravitational 
radiation. 
\bigskip
\bigskip

\ni
{\lbf Appendix: Special spin frames and asymptotic symmetries of ${\cal 
I}^+$}\par
\bigskip
\ni
On the bundle ${\bf S}^A({\cal S})$ of unprimed spinors over ${\cal 
S}$ two connections can be introduced in a natural way by means of the 
spacetime connection. The first is the simple projection to ${\cal S}$ 
of the spacetime covariant derivative operator, $\Delta_e:=\Pi_e^f
\nabla_f$, where the orthogonal projection to ${\cal S}$ can be given 
e.g. by $\Pi^a_b=-m^a\bar m_b-\bar m^am_b$. The other is the spinor 
form of the covariant derivative operator $\delta_e$ defined on any 
spacetime vector field $X^a$ by $\delta_eX^a:=\Pi^a_b\Delta_e(\Pi^b_c
X^c)+(\delta^a_b-\Pi^a_b)\Delta_e((\delta^b_c-\Pi^b_c)X^c)$. The 
difference of these two connections is essentially the extrinsic 
curvature of ${\cal S}$ in the spacetime. (For the details see e.g. 
[50,51].) 

Representing the spinor bundle by the line bundles $E(p,q)$, $p-q\in
{\bf Z}$, the derivative operator $\delta_e$ can be represented by the 
edth and edth-prime operators ${\edth}$ and ${\edth}'$ [52]. 
Explicitly, by the isomorphism ${\bf S}_A({\cal S})\approx E(1,0)
\oplus E(-1,0)$, $\lambda_A\approx(\lambda_0,\lambda_1)$, the 
derivatives $m^e\delta_e\lambda_A$ and $\bar m^e\delta_e\lambda_A$ 
are represented by the cross sections $({\edth}\lambda_0,{\edth}
\lambda_1)$ and $({\edth}'\lambda_0,{\edth}'\lambda_1)$, respectively. 
Denoting e.g. ${\edth}$ acting on the cross sections of $E(p,q)$ by 
${\edth}_{(p,q)}$, on topological 2-spheres $\dim\ker{\edth}_{(p,p+n)}
=\dim\ker{\edth}'_{(p+n,p)}=0$ for any $p\in{\bf R}$ and $n\in{\bf N}$, 
and $\dim\ker{\edth}_{(p+n,p)}=\dim\ker{\edth}'_{(p,p+n)}=1+2n$ for any 
$p\in{\bf R}$ and $n=0,1,2,...$. (For the corresponding kernels on 
higher genus surfaces see [50].) The irreducible parts of $\Delta_e$ 
and $\delta_e$, respectively, are 

$$\eqalignno{
{\cal T}^-(\lambda_0,\lambda_1):=&m^e\bigl(\Delta_e\lambda_A\bigr)o^A
 ={\edth}\lambda_0+\sigma\lambda_1, \hskip 20pt 
 t^-(\lambda_0,\lambda_1):=m^e\bigl(\delta_e\lambda_A\bigr)o^A={\edth}
 \lambda_0,  &(A.0.1.a,b)\cr
\Delta^+(\lambda_0,\lambda_1):=&\bar m^e\bigl(\Delta_e\lambda_A\bigr)
 o^A={\edth}'\lambda_0+\rho\lambda_1, \hskip 20pt
 \delta^+(\lambda_0,\lambda_1):=\bar m^e\bigl(\delta_e\lambda_A\bigr)
 o^A={\edth}'\lambda_0, &(A.0.2.a,b)\cr
-\Delta^-(\lambda_0,\lambda_1):=&m^e\bigl(\Delta_e\lambda_A\bigr)\iota
 ^A={\edth}\lambda_1+\rho'\lambda_0, \hskip 14pt
 -\delta^-(\lambda_0,\lambda_1):=m^e\bigl(\delta_e\lambda_A\bigr)\iota
 ^A={\edth}\lambda_1, &(A.0.3.a,b)\cr
-{\cal T}^+(\lambda_0,\lambda_1):=&\bar m^e\bigl(\Delta_e\lambda_A\bigr)
 \iota^A={\edth}'\lambda_1+\sigma'\lambda_0, \hskip 14pt
 -t^+(\lambda_0,\lambda_1):=\bar m^e\bigl(\delta_e\lambda_A\bigr)
 \iota^A={\edth}'\lambda_1. &(A.0.4.a,b)\cr}
$$
\ni
In particular, a spinor field $\lambda_A$ is constant on ${\cal S}$ with 
respect to $\Delta_e$ iff $(\lambda_0,\lambda_1)\in{\rm ker}({\cal T}^-
\oplus\Delta^+\oplus\Delta^-\oplus{\cal T}^+)$. As Bramson showed [53], 
such a spinor field does not exist even on large spheres near the future 
or past null infinity in a general asymptotically flat spacetime; and 
for a finite topological 2-sphere, being the boundary of some compact 
spacelike hypersurface $\Sigma$ on which the dominant energy condition 
is satisfied, the existence of a constant spinor field implies that 
$D(\Sigma)$ must have a {\it pp}-wave metric [46,47]. To weaken the 
notion of the $\Delta_e$-constant spinor fields on ${\cal S}$ or on 
the cuts of ${\cal I}^+$, in principle there are six natural 
possibilities. These are represented by the kernel of the first order 
operators (see [47]) 
$\Delta:=\Delta^+\oplus\Delta^-$, 
${\cal H}^-:=\Delta^-\oplus{\cal T}^-$, 
${\cal H}^+:=\Delta^+\oplus{\cal T}^+$, 
${\cal C}^+:=\Delta^+\oplus{\cal T}^-$, 
${\cal C}^-:=\Delta^-\oplus{\cal T}^+$ and 
${\cal T}:={\cal T}^-\oplus{\cal T}^+$. 
Similarly, $\lambda_A$ is constant with respect to $\delta_e$ iff 
$(\lambda_0,\lambda_1)\in{\rm ker}\,(t^-\oplus\delta^+\oplus\delta^-
\oplus t^+)$. However, it is easy to see that the existence of a 
nontrivial $\delta_e$-constant spinor field implies the vanishing of 
the curvature of $\delta_e$, and then, via e.g. the Gauss--Bonnet 
theorem, that the 2-surface must be a torus (see [50]). Thus if we 
want to weaken the condition $\delta_e\lambda_A=0$, we naturally 
arrive at the first order operators 
$\delta:=\delta^+\oplus\delta^-$, 
$h^-:=\delta^-\oplus t^-$, 
$h^+:=\delta^+\oplus t^+$, 
$c^+:=\delta^+\oplus t^-$, 
$c^-:=\delta^-\oplus t^+$ and 
$t:=t^-\oplus t^+$. 
(This $\delta$ should not be confused with the operator $\delta:=
m^a\nabla_a$ of the Newman--Penrose formalism introduced in section 3.) 

For round spheres of area-radius $r$ (i.e. for a metric sphere in a 
spherically symmetric spacetime whose radius $r$ is defined by $4\pi 
r^2:={\rm Area}({\cal S})$) one has $\sigma=0=\sigma'$, the 
convergences $\rho$ and $\rho'$ are constant on ${\cal S}$, and 
${\edth}={1\over r}\,{}_0{\edth}$ and ${\edth}'={1\over r}\,{}_0
{\edth}'$. In particular, in Minkowski spacetime $\rho=-{1\over r}$ 
and $\rho'={1\over2r}$. 
By (3.2-7) and (3.9), asymptotically, near the future null infinity, 
the operators (A.0.1.a-4.a) tend to ${}_\infty{\cal T}^-(\lambda_0,
\lambda_1):={}_0{\edth}\lambda_0$, ${}_\infty\Delta^+(\lambda_0,
\lambda_1):={}_0{\edth}'\lambda_0-\lambda_1$, $-{}_\infty\Delta^-(
\lambda_0,\lambda_1):={}_0{\edth}\lambda_1+{1\over2}\lambda_0$ and 
$-{}_\infty{\cal T}^+(\lambda_0,\lambda_1):={}_0{\edth}'\lambda_1-
\dot{\bar\sigma^0}\lambda_0$, respectively. The analogous ${}_\infty
\delta^\pm$ and ${}_\infty t^\pm$ are just the corresponding unit sphere 
edth and edth-prime operators. In the rest of this appendix we discuss 
the direct sum operators briefly, calculate their kernels explicitly 
for round spheres, and determine how they are related to the asymptotic 
symmetries of the spacetime at the future null infinity. Although at 
the quasi-local level only ${\cal H}^\pm$ yield acceptable spinor 
fields for (2.2) and (2.3) [47], we will see that asymptotically, near 
the future null infinity any of $h^-$, ${\cal H}^-$, ${\cal C}^+$, 
${\cal T}$ and $t$ (and `at' infinity ${}_\infty\Delta$ also) can be 
used to recover the asymptotic symmetries. 
\bigskip

\ni
{\bf A.1 The Dirac--Witten operators $\Delta$ and $\delta$} \par
\bigskip
\ni
By the definitions $\Delta_{A'A}\lambda^A=\bar\iota_{A'}\Delta^+
(\lambda_0,\lambda_1)-\bar o_{A'}\Delta^-(\lambda_0,\lambda_1)$ and 
$\delta_{A'A}\lambda^A=\bar\iota_{A'}\delta^+(\lambda_0,\lambda_1)-
\bar o_{A'}\delta^-(\lambda_0,\lambda_1)$; i.e. $\Delta(\lambda_0,
\lambda_1)=0$ and $\delta(\lambda_0,\lambda_1)=0$ are just the 
Dirac--Witten equations on ${\cal S}$ with respect to $\Delta_e$ and 
$\delta_e$, respectively. They are elliptic operators with vanishing 
index, and hence in the generic case their kernel is 
zero dimensional. In fact, by ${\rm dim}\,{\rm ker}\,{\edth}_{(-1,0)}
={\rm dim}\,{\rm ker}\,{\edth}'_{(1,0)}=0$ one has ${\rm dim}\,{\rm 
ker}\,\delta=0$. On the other hand, there might be exceptional 
2-surfaces, even among the round spheres, for which ${\rm dim}\,{\rm 
ker}\,\Delta$ is not zero. To find these exceptional round 2-spheres 
first take the ${\edth}$-derivative of $\Delta^+(\lambda_0,\lambda_1)
=0$ and the ${\edth}'$-derivative of $\Delta^-(\lambda_0,\lambda_1)
=0$. We obtain ${\edth}{\edth}'\lambda_0=\rho\rho'\lambda_0$ and 
${\edth}'{\edth}\lambda_1=\rho\rho'\lambda_1$, i.e. $\lambda_0$ and 
$\lambda_1$ are eigenfunctions of ${\edth}{\edth}'$ and ${\edth}'
{\edth}$, respectively, with the same eigenvalue $\rho\rho'$. Thus let 
us expand them as series of the $s=\pm{1\over2}$ spin weighted spherical 
harmonics: $\lambda_0=\sum_{j,m}c_0^{j,m}\,{}_{1\over2}Y_{j,m}$ and 
$\lambda_1=\sum_{j,m}c_1^{j,m}\,{}_{-{1\over2}}Y_{j,m}$, where $j={1
\over2},{3\over2},...$ and $m=-j,-j+1,...,j$ (see e.g. [40]). Recalling 
that ${\edth}\,{}_sY_{j,m}=-{1\over\sqrt{2}r}\sqrt{(j+s+1)(j-s)}\,{}
_{s+1}Y_{j,m}$ and ${\edth}'\,{}_sY_{j,m}={1\over\sqrt{2}r}\sqrt{(j-
s+1)(j+s)}\,{}_{s-1}Y_{j,m}$, the second order equations yield that 

$$
2r^2\rho\rho'=-(j+{1\over2})^2,  \hskip20pt j={1\over2},{3\over2},...
\eqno(A.1.1)
$$
\ni
i.e. {\it the Dirac--Witten operator $\Delta$ can have non-trivial 
kernel only for discrete values of $2r^2\rho\rho'$}. Thus for a given, 
allowed $\rho\rho'$ labelled by $j$ one has $\lambda_0=\sum_mc_0^m\,{}
_{1\over2}Y_{j,m}$ and $\lambda_1=\sum_mc_1^m\,{}_{-{1\over2}}Y_{j,m}$. 
Substituting these into the first order equations we finally get $c^m_1
=\sqrt{2}(j+{1\over2})^{-1}r\rho'\,c_0^m$, i.e. 

$$
\lambda^A=\sum_{m=-j}^jc^m_0\Bigl({\sqrt{2}\over j+{1\over2}}r\rho'\,
{}_{-{1\over2}}Y_{j,m}\,o^A-\,{}_{1\over2}Y_{j,m}\,\iota^A\Bigr).
\eqno(A.1.2)
$$
\ni
In particular, for $j={1\over2}$ (A.1.1) is just the condition that the 
round sphere is a metric sphere in Minkowski spacetime, whenever (A.1.2) 
is the combination of the restriction to ${\cal S}$ of the two constant 
spinor fields given explicitly by (3.1). (For the explicit expression of 
${}_sY_{j,m}$, see e.g. [40,54].) However, apart from the Minkowski 
case, (A.1.1) yields {\it negative} Hawking energy $E_H:=({\rm Area}(
{\cal S}_r)/16\pi G^2)^{1\over2}(1+{1\over2\pi}\oint_{{\cal S}_r}\rho
\rho'{\rm d}{\cal S}_r)={r\over2G}(1-(j+{1\over2})^2)$, which is just 
the (holomorphic and anti-holomorphic) Dougan--Mason energy in an 
appropriate basis, because the mass is $m^2:=\varepsilon_{\bA\bB}
\varepsilon_{{\bA}'{\bB}'}P^{{\bA}{\bA}'}P^{{\bB}{\bB}'}={r^2\over4G^2}
(1+2r^2\rho\rho')^2$. Thus, in general, the operators $\Delta$ and 
$\delta$ do not define special spinor fields on round spheres. 

On the other hand, on {\it large spheres} near the future null infinity 
the Dirac--Witten equations tend to ${}_\infty\Delta(\lambda_0,\lambda
_1)=0$, the Dirac--Witten equations on the unit round sphere in Minkowski 
spacetime. But ${}_\infty\Delta(\lambda_0,\lambda_1)=0$ admits ${\cal 
E}^A_0=O^A$ and ${\cal E}^A_1=I^A$, given explicitly by (3.1), as 
independent solutions. Therefore, by the analysis of Section 3, both 
the BMS translation and (up to supertranslations) the BMS boost-rotation 
generators can be recovered from the (exceptional) solutions of the 
Dirac--Witten equations on the cuts of ${\cal I}^+$. However, apart 
from the Minkowski case, these solutions are {\it not} limits of 
solutions of the Dirac--Witten equations on finite, large spheres, 
because the latter's do not exist in general. 
\bigskip

\ni
{\bf A.2 The anti-holomorphy operators ${\cal H}^-$ and $h^-$}\par
\bigskip
\ni
By the definitions $m^e\Delta_e\lambda_A=-o_A\Delta^-(\lambda_0,
\lambda_1)-\iota_A{\cal T}^-(\lambda_0,\lambda_1)$ and $m^e\delta_e
\lambda_A=-o_A\delta^-(\lambda_0,\lambda_1)-\iota_At^-(\lambda_0,
\lambda_1)$; i.e. $\lambda_A$ is anti-holomorphic with respect to 
$\Delta_e$ or $\delta_e$ iff ${\cal H}^-(\lambda_0,\lambda_1)=0$ or 
$h^-(\lambda_0,\lambda_1)=0$, respectively. The symplectic scalar 
product of any two anti-holomorphic spinor fields (with respect to 
either $\Delta_e$ or $\delta_e$) is anti-holomorphic, and hence constant 
on ${\cal S}$. ${\cal H}^-$ and $h^-$ are elliptic, and their index 
is $2(1-g)$, where $g$ is the genus of ${\cal S}$. Therefore, in the 
generic case on topological 2-spheres their kernel is two complex 
dimensional. 

In fact, by ${\rm dim}\,{\rm ker}\,{\edth}_{(-1,0)}=0$ and ${\rm dim}
\,{\rm ker}\,{\edth}_{(1,0)}=2$ the $\delta_e$-anti-holomorphic spinor 
fields have the form $\lambda_A=-\lambda_0\iota_A$, where $\lambda_0
\in{\rm ker}\,{\edth}_{(1,0)}$. Therefore, the space of $\delta
_e$-anti-holomorphic spinor fields does not inherit a natural $SL(2,
{\bf C})$ scalar product from $\varepsilon_{AB}$. However, {\it on 
round spheres} these spinor fields can be normalized with respect to 
each other: For $\alpha_A,\,\tilde\alpha_A\in{\rm ker}\,h^-$ the 
combination $\alpha_0\bar\alpha_{0'}+\tilde\alpha_0\bar{\tilde\alpha}
_{0'}$ is constant on ${\cal S}$, and can be required to be $\sqrt2$. 
In fact, on round spheres for the two independent explicit solutions 
we can choose 

$$
\alpha^0_A={{\rm i}\root4\of{2}\,\zeta\over\sqrt{1+\zeta\bar\zeta}}\,
  \iota_A, \hskip12pt
\alpha^1_A={{\rm i}\root4\of{2}\over\sqrt{1+\zeta\bar\zeta}}\,\iota_A, 
\eqno(A.2.1)
$$
\ni
which are normalized in this sense. The transformations leaving this 
normalization invariant is $SL(2,{\bf C})\times U(1)$. (In fact, the 
normalization condition is only {\it one real} condition, and this 
leaves an unspecified phase in the basis solutions (A.2.1).) The spinor 
components $\alpha^{\bA}_Ao^A$ are proportional to the $s={1\over2}$ 
spin weighted spherical harmonics: They are $-\root4\of{2}\sqrt{2\pi}
{}_{1\over2}Y_{{1\over2},\pm{1\over2}}$, respectively, and their 
normalization with respect to the unit sphere volume form is given by 
$\oint_{\cal S}(\alpha^{\bA}_Ao^A)(\bar\alpha^{{\bB}'}_{B'}\bar o^{B'})
{\rm d}{\cal S}=2\sqrt{2}\pi\,{\rm diag}(1,1)=4\pi\,\sigma^{{\bA}
{\bB}'}_0$. Since {\it on large spheres} the equations for the $\delta
_e$-anti-holomorphic spinor fields are just those on the round unit 
sphere, the contravariant form of the independent $\delta
_e$-anti-holomorphic spinor fields on cuts of ${\cal I}^+$ are given by 
$\alpha^A_{\bA}:=\varepsilon^{AB}\alpha^{\bB}_B\varepsilon_{\bB\bA}$; 
i.e. explicitly by $\alpha^A_0=-{\rm i}\root4\of{2}(1+\zeta\bar\zeta)
^{-{1\over2}}\iota^A$ and $\alpha^A_1={\rm i}\root4\of{2}\zeta(1+\zeta
\bar\zeta)^{-{1\over2}}\iota^A$. Comparing these with the functions 
$\tau_{\bA}$ at the end of the first paragraph in Section 3, one can 
see that {\it $\alpha^A_{\bA}=\tau_{\bA}\iota^A=\tau_{\bA}\hat\iota^A$, 
which are just the leading parts of the spinor constituents of the BMS 
translations at ${\cal I}^+$}. The main part of the anti-self-dual BMS 
rotations can also be expressed as $-o_Ao_B\alpha^A_{({\bA}}\alpha^B
_{{\bB})}$. These $\delta_e$-anti-holomorphic spinor fields are used 
in [55] to find a relationship between the Bondi--Sachs mass at the 
{\it past} null infinity and the area of a marginally trapped surface. 

For the independent $\Delta_e$-anti-holomorphic spinor fields {\it on 
round spheres} we choose 

$$
\nu^0_A=-{{\rm i}\over{\root4\of{2}}}{1\over\sqrt{1+\zeta\bar\zeta}}
 \Bigl(2r\rho'o_A-\sqrt2\zeta\iota_A\Bigr), \hskip 12pt
\nu^1_A={{\rm i}\over{\root4\of{2}}}{1\over\sqrt{1+\zeta\bar\zeta}}
 \Bigl(\bar\zeta o_A+{1\over\sqrt2 r\rho'}\iota_A\Bigr). \eqno(A.2.2)
$$
\ni
These are normalized with respect to the pointwise $SL(2,{\bf C})$ 
scalar product: $\varepsilon^{AB}\nu^{\bA}_A\nu^{\bB}_B=\varepsilon
^{\bA\bB}$. Therefore, a natural spin space structure is inherited on 
the space of $\Delta_e$-anti-holomorphic spinor fields on round 
spheres. The $L_2$ scalar product of the spinor components are $\oint
_{\cal S}\nu^{\bA}_0\bar\nu^{{\bB}'}_{0'}{\rm d}{\cal S}=2\sqrt{2}\pi
\,{\rm diag}(1,(2r\rho')^{-2})$ and $\oint_{\cal S}\nu^{\bA}_1\bar\nu
^{{\bB}'}_{1'}{\rm d}{\cal S}=\sqrt{2}\pi\,{\rm diag}((2r\rho')^2,1)$. 
In Minkowski spacetime $\nu^{\bA}_A$ reduces to the restriction to 
${\cal S}$ of $-{\cal E}^{\bA}_A=\{I_A,-O_A\}$, minus the dual of the 
Cartesian spin frame (3.1). Since the equations for 
the $\Delta_e$-anti-holomorphic spinor fields {\it on large spheres} 
are the ones on the round unit sphere in Minkowski spacetime, {\it 
the spinor constituents of the BMS translations on the cuts of ${\cal 
I}^+$ are given by the components of the $\Delta_e$-anti-holomorphic 
spinor fields via $\tau_{\bA}=o_A{\cal E}^A_{\bA}=o_A\varepsilon^{AB}
\nu^{\bB}_B\varepsilon_{\bB\bA}$, and hence the spinor constituents 
themselves are $\varepsilon^{AB}\nu^{\bB}_B\varepsilon_{\bB\bA}=\tau
_{\bA}\hat\iota^A+O(\Omega)$}. This representation of the BMS 
translations is used in [35]. The expression of the main part of the 
anti-self-dual BMS rotations by the $\Delta_e$-anti-holomorphic 
spinors is then obvious. 
\bigskip

\ni
{\bf A.3 The holomorphy operators ${\cal H}^+$ and $h^+$}\par
\bigskip
\ni
As a simple consequence of the definitions, $\lambda_A$ is holomorphic 
with respect to $\Delta_e$ or to $\delta_e$ iff ${\cal H}^+(\lambda_0,
\lambda_1)=0$ or $h^+(\lambda_0,\lambda_1)=0$, respectively. The 
properties of these differential operators are similar to those of 
${\cal H}^-$ and $h^-$, thus we concentrate only the differences. By 
${\rm dim}\,{\rm ker}\,{\edth}'_{(1,0)}=0$ and ${\rm dim}\,{\rm ker}\,
{\edth}'_{(-1,0)}=2$ the $\delta_e$-holomorphic spinor fields have the 
form $\lambda_A=\lambda_1o_A$, where $\lambda_1\in{\rm ker}\,{\edth}'
_{(-1,0)}$. {\it On round spheres} for the two independent explicit 
solutions we can choose 

$$
\beta^0_A=-{{\rm i}\over\root4\of{2}}{1\over\sqrt{1+\zeta\bar\zeta}}\,
 o_A, \hskip12pt
\beta^1_A={{\rm i}\over\root4\of{2}}{\bar\zeta\over\sqrt{1+\zeta\bar
 \zeta}}\,o_A,\eqno(A.3.1)
$$
\ni
whose components $\beta^{\bA}_A\iota^A$ are proportional to the $s=-
{1\over2}$ spin weighted spherical harmonics: They are $-\root4\of{2}
\sqrt{\pi}\,{}_{-{1\over2}}Y_{{1\over2},\pm{1\over2}}$, respectively. 
Their contravariant form are the spinor constituents of the BMS 
translations on the {\it past} null infinity ${\cal I}^-$, and {\it 
not} on ${\cal I}^+$. {\it On round spheres} 

$$
\mu^0_A={{\rm i}\over{\root4\of{2}}}{1\over\sqrt{1+\zeta\bar\zeta}}
 \Bigl({1\over r\rho}o_A+\zeta\sqrt2 \iota_A\Bigr), \hskip 12pt
\mu^1_A={{\rm i}\over{\root4\of{2}}}{1\over\sqrt{1+\zeta\bar\zeta}}
 \Bigl(\bar\zeta o_A-r\rho\sqrt2 \iota_A\Bigr) \eqno(A.3.2)
$$
\ni
form a normalized spin frame in the space of $\Delta_e$-holomorphic 
spinor fields, which in Minkowski spacetime reduce to $\{I_A,-O_A\}$. 
However, the $\Delta_e$-holomorphic spinor equations {\it on large 
spheres} near the {\it future} null infinity do {\it not} reduce to 
the ones on a round sphere: Because of the $\dot{\bar\sigma^0}$ term 
in ${}_\infty{\cal T}^+$ the solutions of ${}_\infty{\cal H}^+(\lambda
_0,\lambda_1)=0$ are {\it not} the restriction to ${\cal S}$ of the 
constant spinor fields of the Minkowski spacetime. They are constant 
only if $\dot\sigma^0=0$, i.e. in absence of outgoing gravitational 
radiation. 
\bigskip

\ni
{\bf A.4 The operators ${\cal C}^+$ and $c^+$}\par
\bigskip
\ni
By the definitions the equations ${\cal C}^\pm(\lambda_0,\lambda_1)=0$ 
and $c^\pm(\lambda_0,\lambda_1)=0$ can be written into the manifestly 
covariant form $(\Delta_a\lambda_B)\pi^{\pm B}{}_C=0$ and $(\delta_a
\lambda_B)\pi^{\pm B}{}_C=0$, respectively, where $\pi^{+A}{}_B=\iota^A
o_B$ and $\pi^{-A}{}_B=-o^A\iota_B$ are the projections of the spin 
spaces onto the space of the $\pm1$ eigenspinors of the spinor $\gamma
^A{}_B=o^A\iota_B+\iota^Ao_B$, defining a GHP-spin frame independent 
notion of chirality on the spin spaces (see [51]). 
${\cal C}^\pm$ and $c^\pm$ are not elliptic operators. In fact, by 
${\rm dim}\,{\rm ker}\,{\edth}_{(1,0)}=2$ and ${\rm dim}\,{\rm ker}\,
{\edth}'_{(1,0)}=0$ on round spheres the kernel of ${\cal C}^+$ is two 
dimensional only if $\rho\not=0$, otherwise ${\rm dim}\,{\rm ker}\,
{\cal C}^+=\infty$. In particular, $c^+$ is just the $\rho=0$ special 
case of ${\cal C}^+$, whenever the elements of $c^+$ have the form 
$\lambda_A=\lambda_1o_A$ with arbitrary $\lambda_1:{\cal S}\rightarrow
{\bf R}$. 
{\it On round spheres} with non-zero $\rho$ the kernel of ${\cal C}^+$ 
is just the kernel of ${\cal H}^+$; i.e. for the two explicit solutions 
we can choose (A.3.2). However, the equations ${\cal C}^+(\lambda_0,
\lambda_1)=0$ {\it on large spheres} near the {\it future} null infinity 
tend to the asymptotic twistor equations of Bramson [53], which turn 
out to be the equations ${\cal C}^+(\lambda_0,\lambda_1)=0$ {\it on the 
round unit sphere} in Minkowski spacetime, and hence ${\rm ker}\,{}
_\infty{\cal C}^+={\rm ker}\,{}_\infty{\cal H}^+$ still holds. Therefore, 
{\it the $\iota^A$-components of the normalized solutions of the 
asymptotic twistor equation reproduce the components $\tau_{\bA}$ of 
the spinor constituents of the BMS translations on ${\cal I}^+$, and 
the solutions themselves are $\varepsilon^{AB}\mu^{\bB}_B\varepsilon
_{\bB\bA}=\tau_{\bA}\hat\iota^A+O(\Omega)$}. This is one of the most 
popular representation of the BMS translations (see 
[22-24,34,38,40,56]). The expression of the BMS rotations is then 
straightforward.  
\bigskip

\ni
{\bf A.5 The operators ${\cal C}^-$ and $c^-$}\par
\bigskip
\ni
By ${\rm dim}\,{\rm ker}\,{\edth}_{(-1,0)}=0$ and ${\rm dim}\,{\rm ker}
\,{\edth}'_{(-1,0)}=2$ the kernel of ${\cal C}^-$ {\it on round spheres} 
is two dimensional only if $\rho'\not=0$, otherwise ${\rm dim}\,{\rm ker}
\,{\cal C}^-=\infty$. $c^-$ is the $\rho'=0$ special case of ${\cal C}
^-$, and the elements of $c^-$ have the form $\lambda_A=-\lambda_0\iota
_A$ with arbitrary $\lambda_0:{\cal S}\rightarrow{\bf R}$. 
If $\rho'\not=0$ then ${\rm ker}\,{\cal C}^-={\rm ker}\,{\cal H}^-$, 
thus for the two explicit solutions of ${\cal C}^-(\lambda_0,\lambda_1)
=0$ we can choose (A.2.2). However, the equations ${\cal C}^-(\lambda_0,
\lambda_1)=0$ {\it on large spheres} near the {\it future} null infinity 
are {\it not} the equations ${\cal C}^-(\lambda_0,\lambda_1)=0$ on the 
round unit sphere in Minkowski spacetime, just because of the presence 
of the $\dot{\bar\sigma^0}$ term in ${}_\infty{\cal T}^+$. Therefore, 
the solutions of ${\cal C}^-(\lambda_0,\lambda_1)=0$ on large spheres 
can be used to represent the BMS translations on the {\it past} null 
infinity. 
\bigskip

\ni
{\bf A.6 The 2-surface twistor operators ${\cal T}$ and $t$}\par
\bigskip
\ni
The 2-surface twistor equations ${\cal T}(\lambda_0,\lambda_1)=0$ 
can be written into the covariant form ${\cal T}_{E'EA}{}^B\lambda_B
:=\Delta_{E'(E}\lambda_{A)}$ $+{1\over2}\gamma_{EA}\gamma^{BC}\Delta
_{E'B}\lambda_C=0$; i.e. the vanishing of the $\gamma_{AB}$-trace-free 
symmetrized $\Delta_e$-derivative of $\lambda_A$ (see [45]). Similarly, 
$t(\lambda_0,\lambda_1)=0$ is equivalent to $\delta_{E'(E}\lambda_{A)}
+{1\over2}\gamma_{EA}\gamma^{BC}\delta_{E'B}\lambda_C=0$. ${\cal T}$ 
and $t$ are elliptic operators with index $4(1-g)$. {\it On round 
spheres} they coincide, and by ${\rm dim}\,{\rm ker}\,{\edth}_{(1,0)}=
{\rm dim}\,{\rm ker}\,{\edth}'_{(-1,0)}=2$ their kernel is four 
dimensional. Therefore, the general explicit solution of the 2-surface 
twistor equations has the form $\omega_A=a_{\bA}\alpha^{\bA}_A+b_{\bA}
\beta^{\bA}_A$ for complex constants $a_{\bA}$, $b_{\bA}$, where 
$\alpha^{\bA}_A$ and $\beta^{\bA}_A$ are given by (A.2.1) and (A.3.1), 
respectively. Introducing $\pi_{A'}:={\rm i}\,\Delta_{A'A}\omega^A$ and 
the 2-surface twistor ${\tt Z}^\alpha:=(\omega^A,\pi_{A'})$, the usual 
Hermitian pointwise scalar product of any two 2-surface twistors ${\tt 
Z}^\alpha$ and ${\tt Z}'^\alpha$ is constant on ${\cal S}$, and 
hence defines the familiar Hermitian scalar product on the space ${\bf 
T}^\alpha$ of the 2-surface twistors. 
However, the flat spacetime definition of the infinity twistor does not 
define any (global) twistor on ${\bf T}^\alpha$, because $\varepsilon
^{A'B'}\pi_{A'}\pi'_{B'}$ is not constant on ${\cal S}$. Its 
modification, ${\tt I}_{\alpha\beta}{\tt Z}^\alpha{\tt Z}'{}^\beta:=
\varepsilon^{A'B'}\pi_{A'}\pi'_{B'}+{1\over2r^2}(1+2r^2\rho\rho')
\varepsilon_{AB}\omega^A\omega'^B$ is constant on ${\cal S}$, but, 
apart from the Minkowski case $2r^2\rho\rho'=-1$, its rank is four 
and hence it fails to be simple. Thus, even on general round spheres, 
${\tt I}_{\alpha\beta}{\tt Z}^\beta=0$ cannot be used to {\it define} 
`translations' in the twistor space ${\bf T}^\alpha$. 
The primary parts of the 2-surface twistors defined by $t$ and ${\cal 
T}$ coincide, but their secondary parts do not. The 2-surface twistors 
defined by $t$ can be recovered from those of ${\cal T}$ as the $\rho
=0=\rho'$ special case. 

On {\it large spheres} near ${\cal I}^+$ the solutions of the 2-surface 
twistor equations have the structure $\omega_A=a_{\bA}\alpha^{\bA}_A+b
_{\bA}\beta^{\bA}_A+a_{\bA}\gamma^{\bA}o_A=(o_C\varepsilon^{CB}\alpha
^{\bA}_Ba_{\bA})\iota_A+(b_{\bA}\beta^{\bA}_B\iota^B+a_{\bA}\gamma
^{\bA})o_A$, where $\gamma^{\bA}$ are of spin weight $-{1\over2}$ and 
satisfy ${}_0{\edth}'\gamma^{\bA}=\dot{\bar\sigma^0}\alpha^{\bA}_0$. 
Comparing this with the results of subsection {\bf A.2} we find that 
$\omega^A=a_{\bA}\tau^{\bA}\iota^A+(b_{\bA}\beta^{\bA}_1+a_{\bA}\gamma
^{\bA})o^A=a_{\bA}\tau^{\bA}\hat\iota^A+O(\Omega)$; i.e. the primary 
part of the 2-surface twistors ${\tt Z}^\alpha=(\omega^A,{\rm i}\,
\Delta_{A'B}\omega^B)$ are just the spinor constituents of the BMS 
translations. Although the 2-surface twistors on large spheres still 
form a 4 dimensional complex vector space, half of them (parametrized 
by the $b_{\bA}$'s) die off asymptotically. 
In absence of outgoing gravitational radiation ${}_\infty{\cal T}$ 
reduces to the homogeneous ${}_\infty t$, whose solutions are just 
those of ${\cal T}$ and of $t$ on the unit round sphere. Interestingly 
enough, in the conformal approach of the 2-surface twistor equation on 
the cuts of ${\cal I}^+$ (see [12-20,38,40]) the absence of outgoing 
gravitational radiation yields the homogeneous equations ${}_0{\edth}
\lambda_0=0$, ${}_0{\edth}'\lambda_1=0$ only after an appropriate 
supertranslation. 
\bigskip

\ni
{\lbf Acknowledgments}\par
\bigskip
\ni
I am grateful to Carlos Kozameh, Osvaldo Moreschi, J\"org Frauendiener, 
James Nester and Paul Tod for discussions, valuable remarks and 
stimulating questions. This work was partially supported by the 
Hungarian Scientific Research Fund grant OTKA T030374. 
\bigskip

\ni
{\lbf References}\par
\bigskip
\item{[1]} R. Geroch, Asymptotic structure of spacetime, in {\it 
           Asymptotic Structure of Spacetime}, ed. F.P. Esposito and 
           L. Witten, Plenum Press, New York 1977

\item{[2]} T. Regge, C. Teitelboim, Role of surface integrals in the 
           Hamiltonian formulation of general relativity, Ann. Phys. 
	   {\bf 88} 286 (1974)
\item{[3]} R. Beig, N.\'O Murchadha, The Poincar\'e group as the 
           symmetry group of canonical general relativity, Ann. Phys. 
           {\bf 174} 463 (1987)
\item{[4]} X. Zhang, Angular momentum and positive mass theorem, Commun. 
           Math. Phys. {\bf 206} 137 (1999)

\item{[5]} J. Winicour, L. Tamburino, Lorentz covariant gravitational 
           energy-momentum linkages, Phys. Rev. Lett. {\bf 15} 601 
           (1965)
\item{[6]} R.W. Lind, J. Messmer, E.T. Newman, Equations of motion for 
           the sources of asymptotically flat spaces, J. Math. Phys.
           {\bf 13} 1884 (1972) 
\item{[7]} R. Geroch, J. Winicour, Linkages in general relativity, J. 
           Math. Phys. {\bf 22} 803 (1981)

\item{[8]} A. Ashtekar, M. Streubel, Symplectic geometry of radiative 
           modes and conserved quantities at null infinity, Proc. Roy. 
           Soc. Lond. A {\bf 376} 585 (1981)
\item{[9]} A. Ashtekar, J. Winicour, Linkages and Hamiltonians at null 
           infinity, J. Math. Phys. {\bf 23} 2410 (1982)

\item{[10]} C.R. Prior, Angular momentum in general relativity I. 
            Definition and asymptotic behaviour, Proc. Roy. Soc. Lond. 
            A. {\bf 354} 379 (1977)
\item{[11]} M. Streubel, `Conserved' quantities for isolated gravitational 
            systems, Gen. Rel. Grav. {\bf 9} 551 (1978)
\item{[12]} R. Penrose, Quasi-local mass and angular momentum in general 
            relativity, Proc. Roy. Soc. Lond. {\bf A381} 53 (1982)
\item{[13]} O.M. Moreschi, On angular momentum at future null infinity, 
            Class. Quantum Grav. {\bf 3} 503 (1986)
\item{[14]} O.M. Moreschi, Supercentre of mass system at future null 
            infinity, Class. Quantum Grav. {\bf 5} 423 (1988)
\item{[15]} S. Dain, O.M. Moreschi, General existence proof for rest 
            frame systems in asymptotically flat spacetime, Class. 
            Quantum Grav. {\bf 17} 3663 (2000)
\item{[16]} O.M. Moreschi, Unambiguous angular momentum of radiative 
            spacetimes and asymptotic structure in terms of the center 
            of mass system, (unpublished manuscript) (2001)

\item{[17]} T. Dray, M. Streubel, Angular momentum at null infinity, 
            Class. Quantum Grav. {\bf 1} 15 (1984)
\item{[18]} W.T. Shaw, Symplectic geometry of null infinity and 
            two-surface twistors, Class. Quantum Grav. {\bf 1} L33 (1984)
\item{[19]} T. Dray, Momentum flux at null infinity, Class. Quantum Grav. 
            {\bf 2} L7 (1985)
\item{[20]} A.D. Helfer, The angular momentum of gravitational radiation, 
            Phys. Lett. A {\bf 150} 342 (1990)

\item{[21]} P.G. Bergmann, R. Thomson, Spin and angular momentum in 
            general relativity, Phys. Rev. {\bf 89} 400 (1953)
\item{[22]} B.D. Bramson, Relativistic angular momentum for 
            asymptotically flat Einstein--Maxwell manifolds, Proc. Roy. 
            Soc. Lond. A. {\bf 341} 463 (1975)
\item{[23]} B.D. Bramson, The invariance of spin, Proc. Roy. Soc. Lond. 
            A. {\bf 364} 383 (1978)
\item{[24]} B.D. Bramson, {\it Physics in cone space}, in Asymptotic 
            structure of spacetime, Eds.: P. Esposito and L. Witten, 
	    Plenum Press, New York 1976

\item{[25]} J. Katz, D. Lerer, On global conservation laws at null 
            infinity, Class. Quantum Gravity, {\bf 14} 2249 (1997)

\item{[26]} A. Rizzi, Angular momentum in general relativity: A new 
            definition, Phys. Rev. Lett. {\bf 81} 1150 (1998)
\item{[27]} A. Rizzi, Angular momentum in general relativity: The 
            definition at null infinity includes the spatial definition 
            as a special case, Phys. Rev. D. {\bf 63} 104002 (2001) 

\item{[28]} C.-M. Chen, J.M. Nester, Quasi-local quantities for general 
            relativity and other gravity theories, Class. Quantum 
            Grav. {\bf 16} 1279 (1999)

\item{[29]} J.D. Brown, J.W. York, Quasi-local energy and conserved 
            charges derived from the gravitational action, Phys. Rev. 
            D {\bf 47} 1407 (1993)
\item{[30]} J.D. Brown, S.R. Lau, J.W. York, Action and energy of the 
            gravitational fields, gr-qc/0010024 
\item{[31]} S.R. Lau, Lightcone reference for total gravitational 
            energy, Phys. Rev. D.{\bf 60} 104034 (1999)
\item{[32]} J.D. Brown, S.R. Lau, J.W. York, Canonical quasi-local 
            energy and small spheres, Phys. Rev. D {\bf 59} 064028 
            (1999)
\item{[33]} R.J. Epp, Angular momentum and an invariant quasi-local 
            energy in general relativity, Phys. Rev. D {\bf 62} 124018 
            (2000)

\item{[34]} M. Ludvigsen, J.A.G. Vickers, Momentum, angular momentum
            and their quasi-local null surface extensions, J. Phys. A:
            Math. Gen. {\bf 16} 1155 (1983)
\item{[35]} A.J. Dougan, L.J. Mason, Quasilocal mass constructions with 
	     positive energy, Phys. Rev. Lett. {\bf 67} 2119 (1991)
\item{[36]} L.B. Szabados, Quasi-local energy-momentum and two-surface
            characterization of the pp-wave spacetimes, Class. Quantum 
	     Grav. {\bf 13} 1661 (1996)
\item{[37]} L.B. Szabados, On certain quasi-local spin-angular momentum 
            expressions for small spheres, Class. Quantum Grav. {\bf 16} 
            2889 (1999)

\item{[38]} W.T. Shaw, The asymptopia of quasi-local mass and momentum 
            I. General formalism and stationary spacetimes, Class. 
	    Quantum Grav. {\bf 3} 1069 (1986) 
\item{[39]} A.J. Dougan, Quasi-local mass for spheres, Class. Quantum
            Grav. {\bf 9} 2461 (1992)

\item{[40]} R. Penrose, W. Rindler, {\it Spinors and Spacetime}, vols. 1 
            and 2, Cambridge Univ. Press, Cambridge 1982 and 1986

\item{[41]} C. M\o ller, Conservation laws and absolute parallelism in 
            general relativity, Mat. Fis. Skr. Dan. Vid. Selsk. {\bf 1} 
	     No 10 (1961)
\item{[42]} L.B. Szabados, Canonical pseudotensors, Sparling's form 
            and Noether currents, KFKI Report, 1991-29/B
\item{[43]} L.B. Szabados, On canonical pseudotensors, Sparling's form 
            and Noether currents, Class. Quantum Grav. {\bf 9} 2521 
            (1992)
\item{[44]} V.C. de Andrade, L.C.T. Guillen, J.G. Pereira, Gravitational 
            energy-momentum density in teleparallel gravity, Phys. Rev. 
            Lett. {\bf 84} 4533 (2000) 
\item{[45]} L.B. Szabados, Quasi-local energy-momentum and angular 
            momentum in general relativity I.: The covariant Lagrangian 
            approach, (unpublished manuscript) 

\item{[46]} L.B. Szabados, On the positivity of the quasi-local mass,
            Class. Quantum Grav. {\bf 10} 1899 (1993)
\item{[47]} L.B. Szabados, Two dimensional Sen connections and 
            quasi-local energy-momentum, Class. Quantum Grav. {\bf 11}
            1847 (1994)

\item{[48]} E.T. Newman, R. Penrose, New conservation laws for zero-mass 
            fields in asymptotically flat spacetime, Proc. Roy. Soc. Lond. 
	     A {\bf 305} 175 (1968)
\item{[49]} E.T. Newman, K.P. Tod, {\it Asymptotically flat spacetimes}, 
            in General relativity and gravitation, vol 2, Ed. A. Held, 
	    Plenum Press, New York 1980

\item{[50]} J. Frauendiener, L.B. Szabados, The kernel of the edth 
            operators on higher-genus spacelike 2-surfaces, Class. 
            Quantum Grav. {\bf 18} 1003 (2001)
\item{[51]} L.B. Szabados, Two dimensional Sen connections in general 
            relativity, Class. Quantum Grav. {\bf 11} 1833 (1994)
\item{[52]} R. Geroch, A. Held, R. Penrose, A space-time calculus based 
            on pairs of null directions, J. Math. Phys. {\bf 14} 874 (1973)

\item{[53]} B.D. Bramson, The alignment of frames of reference at null 
            infinity for asymptotically flat Einstein--Maxwell manifolds, 
	    Proc. Roy. Soc. Lond. A. {\bf 341} 451 (1975)

\item{[54]} J. Stewart, {\it Advanced general relativity}, Cambridge 
            Univ. Press, Cambridge 1990

\item{[55]} M. Ludvigsen, J.A.G. Vickers, An inequality relating total 
            mass and the area of a trapped surface in general relativity, 
            J. Phys. A.: Math. Gen. {\bf 16} 3349 (1983)
\item{[56]} G.T. Horowitz, K.P. Tod, A relation between local and 
            total energy in general relativity, Commun. Math. Phys. 
            {\bf 85} 429 (1982)
\end